\documentclass[12pt]{article}
\pdfoutput=1
\usepackage{jheppub}

\usepackage{slashed}
\usepackage{color, verbatim}
\usepackage{latexsym}
\usepackage{epsfig}
\usepackage{amsmath,accents}

\usepackage{amssymb}
\usepackage{graphicx}
\usepackage{bm}
\usepackage{stackrel}

\usepackage[font={small}]{caption}

\usepackage{hyperref}
\usepackage{epstopdf}
\definecolor{myred}{rgb}{0.7, 0, 0}
\definecolor{myblue}{rgb}{0, 0, 0.7}
\definecolor{mygreen}{rgb}{0.04, 0.7, 0.5}
\hypersetup{colorlinks,citecolor=myred,linkcolor=myblue,urlcolor=myblue,linktocpage=true}

\makeatletter

\@addtoreset{equation}{section}
\makeatother

\newcommand{\be}{\begin{equation}}
\newcommand{\ee}{\end{equation}}
\newcommand{\bea}{\begin{eqnarray}}
\newcommand{\eea}{\end{eqnarray}}

\newcommand{\diag}{\operatorname{diag}}

\begin{document}

\thispagestyle{empty}

%\begin{comment}
\begin{center}

%\hfill UAB-FT-779

\begin{center}

\vspace{.5cm}

{\Large\sc
On Broad Kaluza-Klein Gluons
\vspace{0.3cm}
}\\

\end{center}

\vspace{1.cm}

\textbf{
Rafel Escribano$^{\,a,\,b}$, Mikel Mendizabal$^{\,c}$, Mariano Quir\'os$^{\,b}$, Emilio Royo$^{\,a,\,b}$
}\\

\vspace{.1cm}
${}^a\!\!$ {\em{Grup de Física Teòrica, Departament de Física, Universitat Autònoma de Barcelona, 08193 Bellaterra (Barcelona), Spain}}

${}^b\!\!$ {\em{Institut de Física d'Altes Energies (IFAE), The Barcelona Institute of Science and Technology, Campus UAB, 08193 Bellaterra (Barcelona), Spain}}

${}^c\!\!$ {\em{Deutsches Elektronen-Synchrotron, 22607 Hamburg, Germany}}

\end{center}

\vspace{0.8cm}

\centerline{\bf Abstract}
\vspace{2 mm}

\begin{quote}
\small
In theories with a warped extra dimension, composite fermions, as e.g.~the right-handed top quark, can be very strongly coupled to Kaluza-Klein (KK) fields. 
In particular, the KK gluons in the presence of such composite fields become very broad resonances, thus remarkably modifying their experimental signatures.
We have computed the pole mass and the pole width of the KK gluon, triggered by its interaction with quarks, 
as well as the prediction for proton-proton cross-sections using the full propagator and compared it with that obtained from the usual Breit-Wigner approximation.
We compare both approaches, along with the existing experimental data from ATLAS and CMS, 
for the $t\bar t$, $t\bar t W$, $t\bar t Z$, $t\bar t H$, and $t t \bar t \bar t$ channels. 
We have found differences between the two approaches of up to about 100\%, highlighting that the effect of broad resonances can be dramatic on present, 
and mainly future, experimental searches. 
The channel $t t \bar t \bar t$ is particularly promising because the size of the cross-section signal is of the same order of magnitude as the 
Standard Model prediction, and future experimental analyses in this channel, especially for broad resonances, 
can shed light on the nature of possible physics beyond the Standard Model.     
\end{quote}

\vfill
 
\newpage

\tableofcontents

\newpage
\section{Introduction}
\label{sec:introduction}
The solution to the problem of the Standard Model (SM) sensitivity to ultra-violet (UV) physics, also known as the hierarchy problem, 
generally implies new physics at the TeV scale.
Two main solutions have been proposed thus far: supersymmetry~\cite{Martin:1997ns}, and a warped extra dimension~\cite{Randall:1999ee} \footnote{The 
latter is dual to a conformal field theory (CFT) or composite Higgs theory.}.
The most strongly coupled and model independent fields in these theories are, respectively, the supersymmetric gluino $\tilde g$ 
(a spin $1/2$ Majorana fermion) and the (first) Kaluza-Klein (KK) massive gluon $G$ (a spin $1$ boson), 
with the same quantum numbers as the gluon, and mass around the TeV. 
Even though the spectra of particles in both theories are very rich, the production of these strongly coupled particles is considered 
as the smoking gun of the corresponding theories.

Elusiveness of experimental data at the LHC renders it relevant to reconsider, on more general grounds, the hierarchy problem. 
While gluinos in supersymmetric theories where $R$-parity is conserved appear in reactions as missing energy, e.g.~$\tilde g\to q \tilde q $, 
where the gluino mass cannot be reconstructed, present experimental searches on KK-gluon production focus on searching for (narrow) bumps 
on the invariant mass of its decay products, e.g.~in $G\to q\bar q$. 

As an attempt to explain the above lack of experimental evidence, theories where the KK gluons constitute a continuum above a mass gap, 
instead of isolated resonances, have recently been proposed~\cite{Stancato:2008mp,
Stancato:2010ay,Falkowski:2008yr,Cabrer:2009we,Englert:2012dq,Englert:2012cb,
Goncalves:2018pkt,Csaki:2018kxb,Lee:2018fxj,Megias:2019vdb,Shirazi:2019bjw,Gao:2019gfw}, 
as well as unparticle models~\cite{Georgi:2007ek}.
In this work, though, we want to explore a more conventional solution: the case of broad KK gluons, with mass $M$ and total width $\Gamma$. 
On one hand, as it is well known, given that KK gluons are generically strongly coupled to heavy SM fermions, the associated decay channels can naturally make them broad resonances. 
On the other hand, while production cross-sections decrease with increasing values of the mass $M$, 
for a fixed value of $M$ experimental bounds tend to deteriorate for larger values of the width $\Gamma$, 
or, in other words, for increasing values of the ratio 
\be
r\equiv \Gamma/M\, ,
\label{eq:ratio}
\ee 
which leads to milder experimental bounds from direct searches.

In this work, we will concentrate in the production of the first KK mode ($G\equiv G^{(1)}$) of gluons, 
as they are the most strongly coupled and model independent extra particles in the theory, 
and consider the effect of broad resonances by computing the pole masses and pole decay widths as the zeros of the inverse renormalized propagator; 
as it will be shown, this effect cannot be neglected for strongly coupled fields and broad resonances. 
We will also consider the minimal and most model independent case where only decay channels $G\to q\bar q$, 
where $q$ are the SM quarks, are considered, as the decay channel $G\to gg$ is forbidden. 

By using  {\sc MadGraph5$_-$aMC}~\cite{Alwall:2014hca}, we will apply our results to present LHC bounds on $t\bar t$ as final state. 
Previous similar studies have concentrated on narrower ($r<0.2$) resonances~\cite{Agashe:2006hk,Lillie:2007yh}, 
or tried to fit the large forward-backward (FB) asymmetry at the Tevatron using decay channels involving 
%KK fermion resonances~\cite{Barcelo:2011fw,Barcelo:2011vk}, 
top-quark partners~\cite{Barcelo:2011fw,Barcelo:2011vk,Dasgupta:2019yjm}, 
or considered the case of KK resonances in the electroweak (model dependent) sector, 
which then allows lepton channels as possible final states~\cite{Liu:2019bua,Jung:2019iii}. 
We will also apply our theoretical results to the study of other final state channels, in particular to $t\bar t W$, $t\bar t Z$, $t\bar t H$, and $t t \bar t \bar t$, 
whose experimental detection is currently work in
progress~\cite{Sirunyan:2019wxt,Aad:2020klt,Sirunyan:2020icl,ATLAS:2019nvo,CMS:2019too,ATLAS:2020cxf}. 

The present work is organized as follows. 
In Sec.~\ref{sec:model} we introduce the class of five-dimensional (5D) models, whose first KK gluon $G$ is considered in this work. 
%Readers that are not interested in this kind of formal developments can readily jump to the next section. 
In Sec.~\ref{sec:pole} we compute the pole mass and pole width of the $G$ resonance by means of one-loop diagrams with propagating quarks, 
and compare the pole approach for the full propagator with the on-shell approach, which motivates the Breit-Wigner (BW) approximation. 
The top-quark pair cross-section production $pp\to t\bar t X$ is studied in Sec.~\ref{sec:top_pair} 
as a function of the renormalized mass for fixed values of the width, and as a function of the width for fixed values of the renormalized mass. 
In both cases we compare our predictions with the available experimental data from ATLAS. 
The results presented in this section attempt to motivate the use of the full propagator in the experimental analysis of broad resonances.
In Secs.~\ref{sec:WZH} and \ref{ppttttCS}, we compare our predictions for the channels $t\bar t W$, $t\bar t Z$ and $t\bar t H$, and $t t \bar t \bar t$, 
as functions of the renormalized mass and width, with the existing experimental data. 
The present uncertainty of the experimental data does not allow to draw strong conclusions, albeit the $t t \bar t \bar t$,
 where the SM prediction is of the same order of magnitude as the contribution from new physics, 
 appears to be the most promising avenue for future experimental and theoretical endeavors. 
 Finally, our conclusions are presented in Sec.~\ref{sec:conclusions}. 

\section{The model}
\label{sec:model}
Our model is based on a 5D theory which contains a warped extra dimension $y$, 
with metric given as $ds^2=e^{-A(y)}\eta_{\mu\nu}dx^\mu dx^\nu-dy^2$ in proper coordinates, where $\eta_{\mu\nu}=\diag(1,-1,-1,-1)$, 
and possessing ultra-violet (UV) and infra-red (IR) branes at $y=0$ and $y=y_1$, respectively. 
There is also a bulk propagating field $\phi$ which must stabilize the brane-to-brane distance and give a mass to the 
(otherwise massless) radion~\cite{Goldberger:1999uk}. 
In fact, one can trade the fifth dimension $y$ by the value of the field $\phi$ \footnote{Note 
that $\phi$ is fixed to values $\phi_0$ and $\phi_1$ by potentials localized in the UV and IR branes, respectively.} 
by using the superpotential $W(\phi)$~\cite{DeWolfe:1999cp}, 
which is a function that characterizes the 5D gravitational metric as 
\be
\frac{dA}{d\phi}=\frac{\kappa^2}{3}\left(\frac{\partial \log W}{\partial \phi}\right)^{-1}\ ,
\ee
and gives rise to the bulk potential 
\be
V(\phi)=\frac{1}{8}\left(\frac{\partial W}{\partial \phi}\right)^2-\frac{\kappa^2}{6}W^2(\phi)\ ,
\ee
where $\kappa^2=1/(2M^3)$ and $M$ is the 5D Planck scale.

The metric originally proposed by Randall and Sundrum (RS)~\cite{Randall:1999ee} was Anti-de Sitter in 5D ($AdS_5$) and, thus, 
characterized by a constant superpotential $W_{RS}=6k/\kappa^2$, where $k$ is a constant of the order of the 4D Planck scale. 
This superpotential leads to the linear behavior $A_{RS}(y)=ky$ characteristic of an $AdS_5$ theory. 
It was soon realized that this theory was in conflict with electroweak precision observables, unless the mass of the first KK mode was $\gtrsim 10$ TeV. 
This fact had an impact on the electroweak sector of the theory, which had to be modified to avoid the electroweak constraints. 
Two classes of solutions were advocated in the literature to prevent electroweak bounds. 

One solution, proposed in Ref.~\cite{Agashe:2003zs}, was to enlarge the electroweak gauge sector in the bulk to $SU(2)_L\otimes SU(2)_R\otimes U(1)_{B-L}$,
which contains a custodial symmetry $SU(2)_V$, and where $SU(2)_R\otimes U(1)_{B-L}\to U(1)_Y$ is achieved on the UV brane by boundary conditions, 
while the whole group is unbroken on the IR brane, where custodial symmetry is intact. 
In this way, the SM gauge bosons $W_L^a,B$ have $(+,+)$ boundary condition on the (UV, IR) branes, 
and thus possess zero modes which correspond with the SM gauge bosons, while $W_R^{1,2},Z_R$ have $(-,+)$ boundary conditions, 
and thus only with massive modes~\footnote{These 
massive modes have been recently used to accommodate the $R_{D^{(\ast)}}$ anomalies~\cite{Carena:2018cow}.}. 
In this case we can consider that the back-reaction of the field $\phi$ on the metric is negligible and then the superpotential $W(\phi)\simeq W_{RS}$. 
In this model, left-handed fermions are in $SU(2)_L$ doublets and right-handed ones are arranged in $SU(2)_R$ doublets while the Higgs 
is a fundamental bidoublet in the bulk. 
Furthermore, in this class of theories there are also (gauge-Higgs unification) models, with extended custodial gauge groups, 
for which the Higgs is the fifth component $A_5$ of an odd gauge field $A_\mu$, which gets a mass from a radiative Coleman-Weinberg mechanism. 
Now, the electroweak sector is a general gauge group $\mathcal G$ broken by boundary conditions to $SU(2)_L\otimes U(1)_Y$ 
at the UV brane and to the subgroup $\mathcal H$, which contains a custodial symmetry, on the IR brane. 
Here the Higgs lies along the fifth dimension of the coset group $\mathcal G/\mathcal H$, 
so that the choice of the groups $\mathcal G$ and $\mathcal H$ defines different models~\cite{Contino:2003ve,Agashe:2004rs,Contino:2006qr}. 
In this general class of models, with an enlarged custodial symmetry in the electroweak sector, 
the color group $SU(3)_c$ is unrelated to the electroweak symmetry breaking and thus the gluon sector is pretty much model independent, 
with KK gluons strongly coupled, as in the RS model, to composite fermions, localized toward the IR brane, as the top quark.

A different proposed solution is to use the stabilizing field $\phi$ in a regime where the back reaction on the gravitational metric is important 
near the IR brane and mild near the UV brane, thus, keeping the properties of RS theories to solve the hierarchy problem. 
In these theories, the back-reaction creates a naked singularity beyond the IR brane and they are dubbed soft-wall (SW) metric 
theories~\cite{Cabrer:2009we,Cabrer:2010si,Cabrer:2011fb,Cabrer:2011vu,Carmona:2011ib,Cabrer:2011qb,Quiros:2013yaa,deBlas:2012qf}. 
Typically, theories with strong back-reactions on the metric are characterized by exponential superpotentials, 
as e.g.~$W_{SW}=6k/\kappa^2\left(1+e^{a \phi}  \right)^b$~\cite{Megias:2015ory} with $a$ and $b$ real parameters, 
and a bulk gauge group $SU(3)_c\otimes SU(2)_L\otimes U(1)_Y$, which is just the SM gauge group. 
In this class of models, we can understand the improvement from the electroweak constraints by the fact that the \textit{physical} Higgs boson profile, 
$\tilde h(y)=h(0)e^{\alpha k y-A(y)}$ with $\alpha>2$ to solve the hierarchy problem, unlike the case of the RS metric, 
is peaked away from the IR brane~\cite{Carmona:2011ib}, while the KK modes are localized toward the IR brane, 
which makes their overlap small enough to cope with electroweak constraints for a choice of the model parameters. 
This analysis was performed in Refs.~\cite{Megias:2016bde,Megias:2017ove}. 
Again, in this class of theories as in custodial models, the color group is $SU(3)_c$ and, thus, KK gluons are very much model independent. 

In all previous models, the SM fermions are realized as chiral zero modes of 5D fermions, 
with localization along the fifth dimension determined by a 5D Dirac mass, conveniently chosen as $M_{f_{L,R}}=\mp c_{f_{L,R}}W(\phi)$, 
where the upper (lower) sign applies for a fermion with left-handed (right-handed) zero mode. With this convention, 
the fermion zero modes are localized near the UV (IR) brane for $c_{f_{L,R}}>1/2$ ($c_{f_{L,R}}<1/2$). 
Specifically, their wave-functions in the fifth dimension are given by
\be
\Psi^{(0)}_{f_{L,R}}(x,y)=f_{L,R}(y) \psi_{L,R}(x)\ ,\qquad f_{L,R}(y)=\frac{e^{(2-c_{L,R})A(y)}}{\left[\int dy e^{(1-2 c_{L,R})A}\right]^{1/2}}\ ,
\ee
where $\psi_{L,R}(x)$ are the 4D wave functions. 
Moreover, after interaction with the Higgs field, the 4D mass of the fermion $f$ is determined by the constants $c_{f_L}$ and $c_{f_R}$, 
which in turn determine the overlapping integral with the Higgs wave function, along with the 5D Yukawa couplings. 
In this way, heavy (light) fermions are localized toward the IR (UV) brane and thus have constants $c_f<1/2$ $(c_f>1/2)$.

Likewise, the coupling of the quark $q_{L,R}$ with the KK gluon $G^{(n)}$, with normalized profile $f_{G^{(n)}}$, is given by~\footnote{Notice 
that, due to the different localization of quarks $q_L$ and $q_R$, their interaction with the KK modes $G^{(n)}$ is not vector-like, 
as it happens with its zero mode, the SM gluon $g\equiv G^{(0)}$.}
\be
\label{gGqq}
g_{G^{(n)}q\bar q}=g_s g_q^n(c_q)\ ,\qquad g_q^n(c_q)\equiv \int e^{-3A}q^2(y) f_{G^{(n)}}(y)dy\ ,
\ee
where $g_s$ is the strong coupling constant of $SU(3)_c$. As the KK modes $G^{(n)}$ are strongly localized toward the IR brane,
it turns out that heavy quarks ($c_q<1/2$) can be strongly coupled to $G^{(n)}$, while light quarks ($c_q>1/2$) are weakly coupled. 

\begin{figure}[htb]
\centering
\includegraphics[width=10cm]{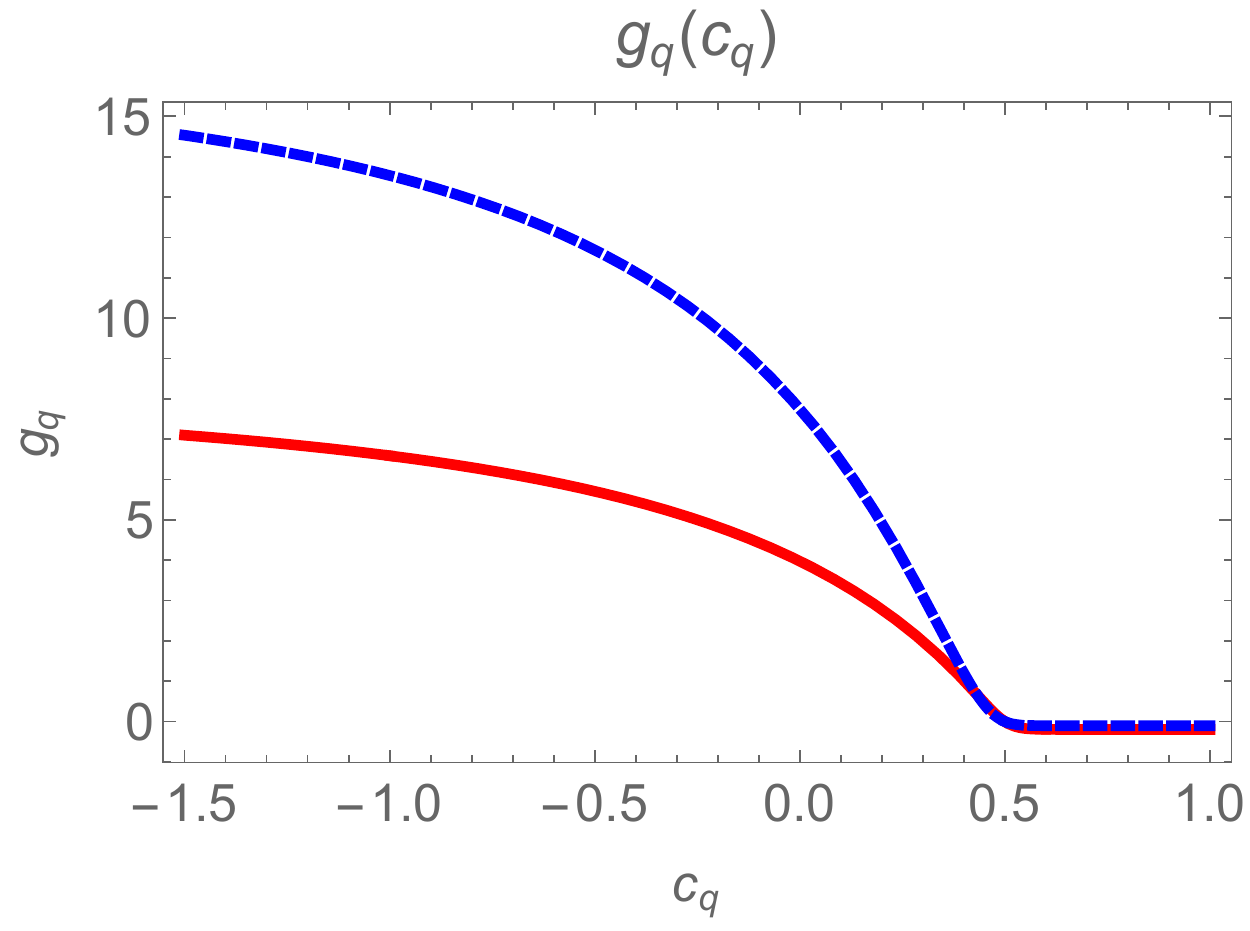} 
\caption{\it Coupling of the first gluon KK mode $G$ with the quark $q$ as a function of its localizing parameter $c_q$ for two models: 
the lower (solid) curve corresponds to the RS metric and the upper (dashed) curve to the SW metric for the choice of parameters $a=0.15$ and $b=2$.}
\label{fig:coupling}
\end{figure}

In the present work, the focus revolves around the couplings of the first KK mode of the gluon (the lightest gluon KK mode) 
with quarks $q_{L,R}$. For illustrative purposes, the coupling $g_q\equiv g_q^1(c_q)$ is shown in Fig.~\ref{fig:coupling} 
as a function of the localizing parameter $c_q$
for the two different class of models considered. 
The lower (solid) curve corresponds to the RS metric and its limiting value for $c_q<1/2,\ |c_q|\gg 1$, is $g_q=\sqrt{2ky_1}\simeq 8.4$. 
In the other limit, when $c_q>1/2,\ |c_q|\gg 1$, the limiting behavior is $g_q\simeq -0.2$.
The upper (dashed) curve corresponds to the considered class of SW models for particular values of the parameters $a=0.15$, $b=2$ 
for which the electroweak oblique observables $S$ and $T$ are minimized~\cite{Megias:2016bde,Megias:2017ove}. 
In this case, one gets the behavior $g_q\simeq -0.2$ for UV localized quarks, while for quarks localized toward the IR we see that $g_q\simeq 15$. 
Needless to say, coupling values $\alpha_q=g_q^2/4\pi\gtrsim 4\pi$ will give rise to non-perturbative behaviors and will make perturbative calculations 
absolutely untrustworthy. 
The plots in Fig.~\ref{fig:coupling} illustrate two sort of extreme examples of 5D theories with a warped extra dimension. 
We expect that for arbitrary theories, the values of the coupling $g_q$ will fall in the region of values that appear in Fig.~\ref{fig:coupling}.

In this work, we consider the right-handed top $t_R$ as the most strongly coupled fermion so that $g_{t_R}$ is kept as a free parameter. 
%Likewise the right-handed bottom $b_R$ will be considered as strongly coupled, and $g_{b_R}$ a free parameter, in view of the $A^b_{FB}$ anomaly 
%which translate into the SM departure for the bottom coupling to the $Z$ gauge boson: 
%$\delta g_{Zb_R\bar b_R}=(2.30\pm 0.88)\times 10^{-2}$~\cite{Falkowski:2017pss}. 
On the other hand, in view of the strong constraints on the coupling of the left-handed bottom to the $Z$ boson, 
$\delta g_{Zb_L\bar b_L}=(0.33\pm 0.17)\times 10^{-2}$~\cite{Falkowski:2017pss}, 
and given that $g_{t_L}=g_{b_L}$ due to $(t_L,b_L)^T$ being part of an $SU(2)_L$ doublet, we will fix $g_{t_L}=g_{b_L}=g_{b_R}=1$. 
In the absence of a UV theory for flavor, fixing the mass of the third generation quarks can be easily done with the help of 5D Yukawa couplings. 
Finally, we will localize the quarks $q_{L}$ and $q_R$ of the first and second generation $q=u,d,c,s$ near the UV brane, 
such that $g_{q_{L,R}}\simeq -0.2$. 

In the next section, we will compute the total decay rate $\Gamma=\sum_q\Gamma(G\to q\bar q)$ and we will see that it depends on the coupling $g_{t_R}$, 
so that we will trade the free parameter $g_{t_R}$ by the total width $\Gamma$. 
Moreover, as the coupling of $G$ with two gluons vanishes, by orthonormality of the wave functions, 
quarks are the only decay channels that we will consider in this work.

\section{Mass and width of the KK gluon}
\label{sec:pole}
In this section, we first calculate
the heavy KK-gluon vacuum polarization (VP) at one-loop,
associated to the effect of virtual quark-antiquark pairs of given chirality,
and, next, resum its effect into the KK-gluon propagator.
After that, we renormalize the propagator following two different approaches:
the \emph{pole approach},
where the renormalized VP function remains exact,
and the \emph{on-shell approach},
where this function is expanded around the KK-gluon mass,
and thus approximated by a Breit-Wigner function.
Finally, we explain how to calculate the corresponding mass and width
of the KK-gluon resonance in each approach and compare both sets of parameters.
%Finally, we introduce an effective parametrization
%by which a Breit-Wigner approximation reproduces the behaviour of
%the exact propagator at the peak and near the resonance.

\subsection{Vacuum polarization and renormalization}
The one-loop contribution from quarks $q=\left(\begin{array}{c}q_L \\ q_R\end{array}\right)$ 
%of given chirality,
%$q_\chi$ with $\chi=L,R$,
to the KK-gluon VP, $\Pi^{ab}_{\mu\nu}(q^2)$, where $a,b=1,\dots,8$ are indices in the $SU(3)$ Lie algebra, 
%calculated using dimension regularization, 
is written as
\begin{equation}
\label{KKgluon1looptensor}
    i\Pi^{ab}_{\mu\nu}(q^2)=i\delta^{ab}
    \left[g_{\mu\nu}q^2\Pi(q^2)-q_\mu q_\nu\Delta(q^2)\right]\ ,
\end{equation}
with $\Pi(q^2)=\bar\Pi(q^2)+\delta M^2/q^2$ and
\begin{equation}
\label{KKgluon1loopcorr}
\begin{array}{rcl}
    \bar\Pi(q^2)
    &=&\Delta(q^2)+\dfrac{g_s^2}{4\pi^2}\sum_q g_{q_A}^2\dfrac{m_q^2}{q^2}
    \left[2+F(q^2)\right]\ ,\\[3ex]
    \Delta(q^2)&=&
    -\dfrac{g_s^2}{24\pi^2}\sum_q (g_{q_V}^2+g_{q_A}^2)
    \left[\dfrac{2}{\varepsilon}-\gamma_E+\ln 4\pi+\ln\dfrac{\mu^2}{m_q^2}
    \right.\\[3ex]
    &&\left.+\dfrac{5}{3}\left(1+\dfrac{12m_q^2}{5q^2}\right)
    +\left(1+\dfrac{2m_q^2}{q^2}\right)F(q^2)\right]\ ,\\[3ex]
    \delta M^2&=&\dfrac{g_s^2}{4\pi^2}\sum_q g_{q_A}^2 m_q^2 
    \left(\dfrac{2}{\varepsilon}-\gamma_E+
    \ln 4\pi+\ln\dfrac{\mu^2}{m_q^2}\right)\ ,
\end{array}
\end{equation}
where $q^2$ is the squared KK-gluon momentum,
$g_s$ is the QCD coupling constant with $\alpha_s=g_s^2/4\pi$,
$g_{q_{V,A}}\equiv (g_{q_L}\pm g_{q_R})/2$
are the vector and axial KK-gluon--quark couplings, in units of $g_s$,
defined in terms of the corresponding left- and right-handed ones,
$\varepsilon$ the regulator of the ultraviolet divergence,
$\gamma_E$ the Euler-Mascheroni constant,
$\mu$ the arbitrary scale which appears in dimensional regularization,
$m_q$ the quark mass, and $F(q^2)$ the loop associated function
given by
\begin{equation}
\label{KKgluon1loopfunction}
F(q^2)=\left\{
\begin{array}{ll}
    \sqrt{1-\dfrac{4m_q^2}{q^2}}
    \left(\ln{\dfrac{1-\sqrt{1-\dfrac{4m_q^2}{q^2}}}
    {1+\sqrt{1-\dfrac{4m_q^2}{q^2}}}}+i\pi\right)    &
    \quad\mbox{for}\quad q^2\ge 4m_q^2\\[8ex]
    -2\sqrt{\dfrac{4m_q^2}{q^2}-1}
    \arctan{\dfrac{1}{\sqrt{\dfrac{4m_q^2}{q^2}-1}}} &
    \quad\mbox{for}\quad q^2<4m_q^2\ .
\end{array}
\right.
\end{equation}

The correction of the former contribution to the KK-gluon propagator 
after resummation is given by
\begin{equation}
\label{KKgluonprop}
\begin{array}{rcl}
iG_{\mu\nu}^{ab}&=&iG_{\mu\nu}^{(0)ab}
+iG_{\mu\alpha}^{(0)ac}
i\Pi^{\alpha\beta}_{cd}(q^2)
iG_{\beta\nu}^{(0)db}+\cdots\\[2ex]
&=&i\delta^{ab}\left[\dfrac{-g_{\mu\nu}+\frac{q_\mu q_\nu}{q^2}}{q^2-M_0^2}
\left(1+\dfrac{q^2\Pi(q^2)}{q^2-M_0^2}\right)
+\dfrac{q_\mu q_\nu}{q^2}\dfrac{1}{M_0^2}\ \mbox{terms}+\cdots\right]\\[4ex]
&=&i\delta^{ab}
\left(\dfrac{-g_{\mu\nu}+\frac{q_\mu q_\nu}{q^2}}{q^2-M_0^2-q^2\Pi(q^2)}
+\dfrac{q_\mu q_\nu}{q^2}\dfrac{1}{M_0^2}\ \mbox{terms}\right)\ ,
\end{array}
\end{equation}
where
$iG_{\mu\nu}^{(0)ab}=i\delta^{ab}
\frac{-g_{\mu\nu}+\frac{q_\mu q_\nu}{M_0^2}}{q^2-M_0^2}$
is the propagator at zeroth order in the unitary gauge and $M_0$ the KK-gluon bare mass.
The $q_\mu q_\nu$ terms give rise to suppressed 
contributions in processes where
the external particles have small masses compared to the
KK-gluon mass, as is the case in the present work.

%\subsection{Renormalization}
The correction of the KK-gluon propagator is divergent and must be renormalized.
In order to regularize $\Pi(q^2)$ in 
Eq.~(\ref{KKgluon1loopcorr}),
one can define $\widehat\Pi(q^2)\equiv\bar\Pi(q^2)-\Re\bar\Pi(q_0^2)$
and thus
\begin{equation}
\label{Mscheme}
\begin{array}{rcl}
q^2-M_0^2-q^2\Pi(q^2)&=&
(q^2-M^2)\left[1-\Re\bar\Pi(q_0^2)\right]-q^2\widehat\Pi(q^2)\\[2ex]
&=&\left[1-\Re\bar\Pi(q_0^2)\right]\left[q^2-M^2-q^2\widehat\Pi(q^2)\right]
+{\cal O}(g^4)\ ,
\end{array}
\end{equation}
where higher powers of the coupling constant are
neglected~\footnote{In the rest of this section,
$g\equiv g_s$ is used for notational simplicity.}.
The parameter $M$ is an observable finite mass defined from
$M_0^2+\delta M^2\equiv M^2\left[1-\Re\bar\Pi(q_0^2)\right]$,
where $q_0$ is some chosen subtraction point used as the renormalization
scale~\footnote{The 
values of $M$ and $g$ depend on the chosen renormalization scale.
For two different subtraction points
$q_{0,i}$, with respective renormalized mass $M_i$ ($i=1,2$), we have 
$M_2^2/M_1^2=1+\Re\bar\Pi(q_{0,2}^2)-\Re\bar\Pi(q_{0,1}^2)+{\cal O}(g^4)$.
The same goes for $g$.}.
The prefactor $1-\Re\bar\Pi(q_0^2)$ in front of the renormalized KK-gluon
propagator is used for the renormalization of the coupling constant
$g_0^2\equiv g^2\left[1-\Re\bar\Pi(q_0^2)\right]$.

Bearing in mind that the KK-gluon propagator is attached to
two vector minus axial-vector currents weighted by the coupling constants,
the net effect of renormalization for its transverse part is~\footnote{The
terms proportional to $q_\mu q_\nu/q^2$ in the longitudinal part of the
propagator are also renormalized:
$g_0^2/M_0^2\rightarrow g^2/M^2+{\cal O}(g^4)$.}
\begin{equation}
\label{KKgluonpropR}
\dfrac{g_0^2\left(-g_{\mu\nu}+\frac{q_\mu q_\nu}{q^2}\right)}
{q^2-M_0^2-q^2\Pi(q^2)}
\longrightarrow
\dfrac{g^2\left(-g_{\mu\nu}+\frac{q_\mu q_\nu}{q^2}\right)}
{q^2-M^2-q^2\widehat\Pi(q^2)}
+{\cal O}(g^6)\ ,
\end{equation}
which from this point onwards is considered as the propagator to be used
in our calculations.

\subsection{Pole approach}
\label{poleapp}
Within the so-called \emph{pole approach} framework \cite{Willenbrock:1991hu,Bhattacharya:1991gr,Escribano:2002iv},
the renormalized propagator in Eq.~(\ref{KKgluonpropR}) does not 
have a pole at $q^2=M^2$.
In the pole approach, the expression $D(q^2)=q^2-M^2-q^2\widehat\Pi(q^2)$
is separated into its real and imaginary parts and is defined in the 
first Riemann sheet as
$D_{\rm I}(s)=s-M^2
-s[\Re\bar\Pi_+(s)-\Re\bar\Pi(q_0^2)]-i s\Im\bar\Pi_+(s)$,
with $s=q^2$ and $\bar\Pi_+(s)\equiv\bar\Pi(s+i\epsilon)$
($\epsilon$ is 
an infinitesimal positive definite parameter).
Due to the presence of the quark threshold~\footnote{In
the following, we consider the approximation that all quarks,
excluding the top quark, are massless.
This approximation is valid since $M\gg m_q$ for $q=u, d, s, c, b$.
Due to this simplification, 
there is effectively only one quark threshold and it is located at $4m_t^2$.},
the complex function $\bar\Pi(s)$ is analytical everywhere in the 
complex plane with exception of a branch cut on the real $s$-axis starting
at $4m_t^2$ up to $+\infty$, the so-called unitary cut, 
which is associated to the bi-valued nature of the square root function.
Accordingly, in addition to the physical sheet (the first Riemann sheet), 
there is an unphysical sheet (the second Riemann sheet)
where the poles associated to resonances are found.
The discontinuity across the cut is given by 
\begin{equation}
\label{Disc}
\begin{array}{rcl}
\textrm{Disc}\bar\Pi(s)&=&\bar\Pi(s+i\epsilon)-\bar\Pi(s-i\epsilon)\\[1ex]
&=&\bar\Pi(s+i\epsilon)-\bar\Pi^\ast(s+i\epsilon)
=2i\Im\bar\Pi_+(s)\ ,
\end{array}
\end{equation}
where in the second equality the Schwarz reflection principle,
$\bar\Pi^\ast(s+i\epsilon)=\bar\Pi(s-i\epsilon)$, has been used
because the function is purely real below the threshold.
The analytical continuation from the first to the second Riemann sheet
is given by
\begin{equation}
\label{IIRS}
\begin{array}{rcl}
\bar\Pi_{\rm II}(s+i\epsilon)&=&\bar\Pi(s-i\epsilon)
=\bar\Pi(s+i\epsilon)-\textrm{Disc}\bar\Pi(s)\\[1ex]
&=&\Re\bar\Pi_+(s)-i\Im\bar\Pi_+(s)\equiv\bar\Pi_-(s)\ .
\end{array}
\end{equation}

In summary, the final form of the propagator's denominator $D(s)$
from which one ought to look for poles in the second Riemann sheet is
\begin{equation}
\label{propIIRS}
D_{\rm II}(s)=s-M^2
-s[\Re\bar\Pi_+(s)-\Re\bar\Pi(q_0^2)]+i s\Im\bar\Pi_+(s)\ .
\end{equation}
Since the function $\bar\Pi(s)$ also contains an implicit dependence
on the modulus of the quark three-momentum, $p(s)=\sqrt{s-4m^2}/2$,
the convention for defining the two different Riemann sheets is
$\textrm{(I,II)}\equiv (+,-):(\Im p>0,\Im p<0)$
for the first and second sheets, respectively.
In practice, one can move from one sheet to the other by replacing 
$p(s)\to -p(s)$ in $D(s)$ and, therefore,
$D_{\rm I}(s)\equiv D[s,p(s)]$ and $D_{\rm II}(s)\equiv D[s,-p(s)]$, 
as can be seen by inspecting Eq.~(\ref{KKgluon1loopcorr}).

Once the parametric values of the KK-gluon mass $M$
and the KK-gluon--quark couplings
(or $g_{t_R}$ in this simplified analysis~\footnote{As stated before,
all quark chiral couplings excluding the top right one
are fixed to predetermined values.
In this way, the analysis is performed just in terms of
$M$ and $g_{t_R}$.}) are given, 
or in the future possibly extracted from LHC experimental data,
the true resonance parameters, the so-called \emph{pole} parameters,
which are model- and process-independent,
are obtained from the solution of the pole equation 
$D_{\rm II}(s_{\rm p})=0$
with $s_{\rm p}\equiv M_{\rm p}^2-i M_{\rm p}\Gamma_{\rm p}$,
where $M_{\rm p}$ and $\Gamma_{\rm p}$
are the \emph{pole mass} and the \emph{pole width}
of the resonance.
If for real $s$, $\bar\Pi_+(s)=\bar\Re\Pi_+(s)+i\Im\bar\Pi_+(s)$,
then for arbitrary complex $s$,
$\bar\Pi_+(s)=\Re\Re\bar\Pi_+(s)-\Im\Im\bar\Pi_+(s)
+i[\Im\Re\bar\Pi_+(s)+\Re\Im\bar\Pi_+(s)]$,
that is,
\begin{equation}
\label{propIIRSsp}
\begin{array}{rcl}
D_{\rm II}(s_{\rm p})&=&s_{\rm p}-M^2
-s_{\rm p}[\Re\Re\bar\Pi_+(s_{\rm p})-\Im\Im\bar\Pi_
+(s_{\rm p})-\Re\bar\Pi(q_0^2)]\\[2ex]
&&-i s_{\rm p}[\Im\Re\bar\Pi_+(s_{\rm p})+\Re\Im\bar\Pi_+(s_{\rm p})]=0\ .
\end{array}
\end{equation}

\subsection{On-shell approach}
An alternative framework, the so-called \emph{on-shell approach} \cite{Willenbrock:1991hu,Bhattacharya:1991gr,Escribano:2002iv},
is based on an expansion of $\bar\Pi(q^2)$ around $q^2=M^2$
allowing the full propagator to have a pole at $q^2=M^2$
with some residue (field-strength renormalization) ${\cal Z}$.
Taking only the real part of the propagator's denominator~\footnote{In
this approach, the subtraction point will be chosen to be $q_0=M$.},
\begin{equation}
\label{OSscheme}
\begin{array}{rcl}
q^2-M_0^2-q^2\Re\Pi(q^2)&=&
q^2-(M_0^2+\delta M^2)-q^2\Re\bar\Pi(q^2)\\[2ex]
&=&(q^2-M^2)\left[1-\Re\bar\Pi(M^2)-q^2\Re\bar\Pi^\prime(M^2)\right]+
{\cal O}(q^2-M^2)^2\\[2ex]
&\equiv&(q^2-M^2){\cal Z}^{-1}+\mathcal O(q^2-M^2)^2\ ,
\end{array}
\end{equation}
where
$\Re\bar\Pi^\prime(M^2)\equiv
\left.\frac{d\Re\bar\Pi(q^2)}{dq^2}\right|_{q^2=M^2}$
and ${\cal Z}^{-1}=1-\Re\bar\Pi(M^2)-M^2\Re\bar\Pi^\prime(M^2)$.
When the imaginary part of $\Pi(q^2)$ is included,
the final form of the relevant part of the propagator is written as
\begin{equation}
\label{KKgluonpropROS}
\begin{array}{rcl}
\dfrac{g_0^2\left(-g_{\mu\nu}+\frac{q_\mu q_\nu}{q^2}\right)}
{q^2-M_0^2-q^2\Pi(q^2)}
&\longrightarrow&
\dfrac{g^2\left[1-\Re\bar\Pi(M^2)\right]{\cal Z}}
{q^2-M^2-i{\cal Z}q^2\Im\bar\Pi(q^2)}
\left(-g_{\mu\nu}+\dfrac{q_\mu q_\nu}{q^2}\right)\\[4ex]
&&=\dfrac{g^2\left[1+M^2\Re\bar\Pi^\prime(M^2)\right]}
{q^2-M^2-i q^2\Im\bar\Pi(q^2)}
\left(-g_{\mu\nu}+\dfrac{q_\mu q_\nu}{q^2}\right)
+{\cal O}(g^6)\ .
\end{array}
\end{equation}
One can easily check that the final expressions for the relevant part
of the propagator appearing in
Eq.~(\ref{KKgluonpropR}), with $q_0=M$, and Eq.~(\ref{KKgluonpropROS})
are equivalent at ${\cal O}(q^2-M^2)^2$.

In the on-shell approach, 
the expression $q^2-M^2-i{\cal Z}q^2\Im\bar\Pi(q^2)$
closely resembles the relativistic Breit-Wigner (BW) formula
$q^2-M^2+i M\Gamma$.
If $q^2\Im\bar\Pi(q^2)$ is small, so that the resonance is narrow,
one can approximate $q^2\Im\bar\Pi(q^2)$ as $M^2\Im\bar\Pi(M^2)$
over the width of the resonance.
Then, one has precisely the BW form with the identification
\begin{equation}
\label{BWwidth}
\Gamma=-{\cal Z}M\Im\bar\Pi(M^2)=-M\Im\bar\Pi(M^2)+{\cal O}(g^4)\ .
\end{equation}
For a broad resonance, as it may well be the case for the heavy KK gluon,
the full energy dependence of $s\Im\bar\Pi(s)$
must be taken into account though.

\subsection{Comparison between on-shell and pole approaches}
\begin{figure}[htb]
\centering
\includegraphics[width=7.5cm]{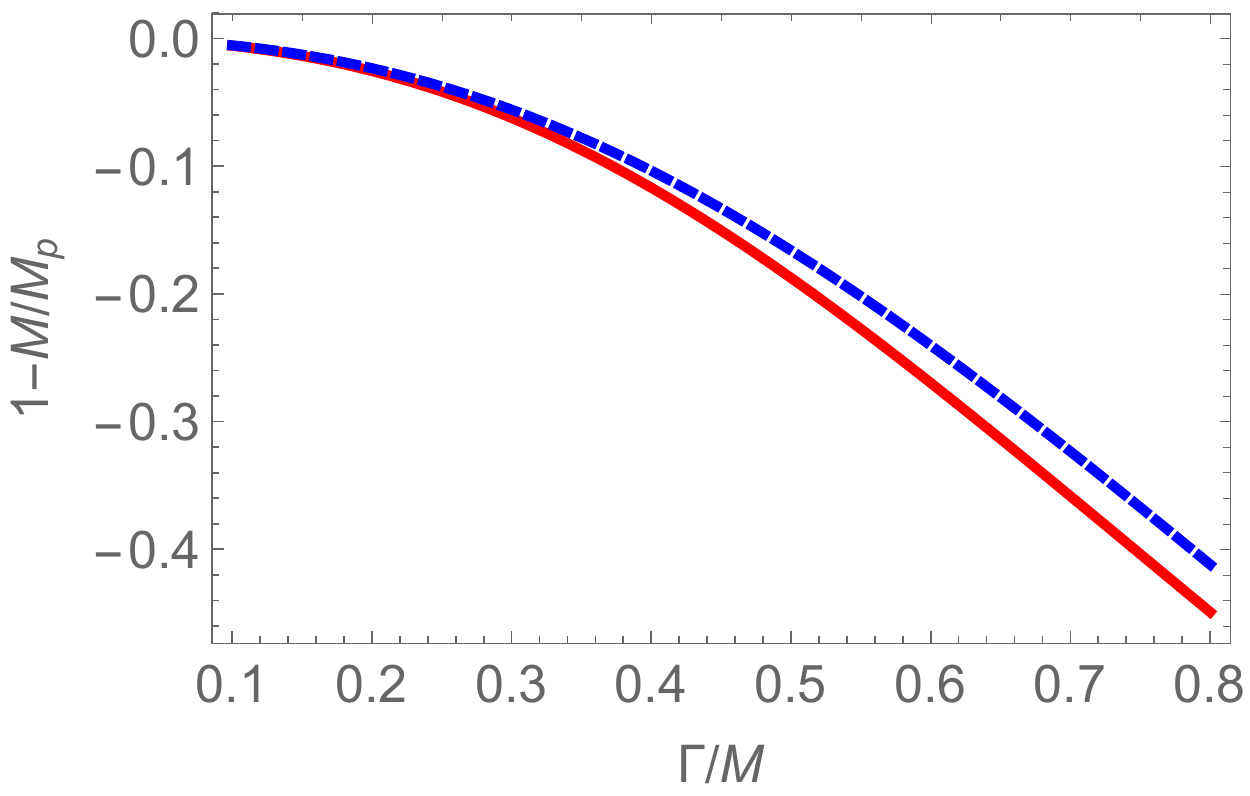}\quad
\includegraphics[width=7.5cm]{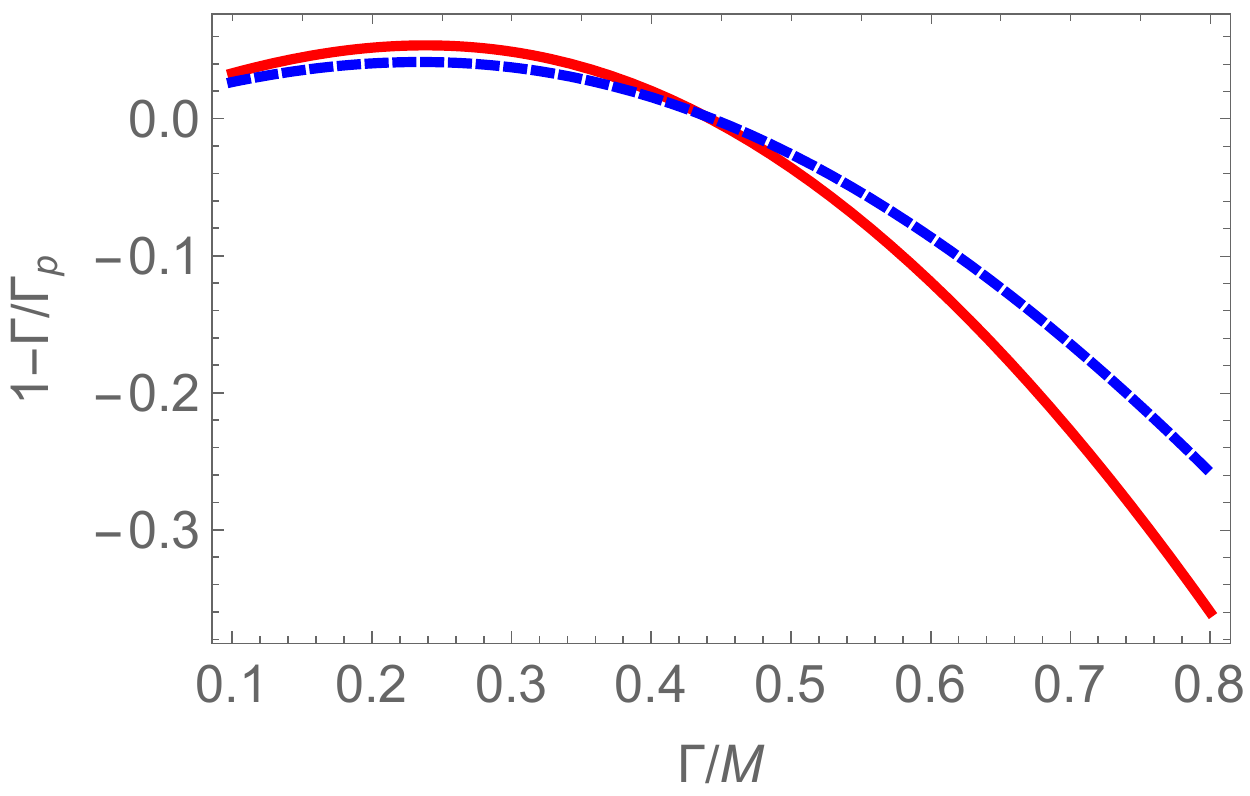}
\caption{\it Plot of $1-M/M_{\rm p}$ \emph{(left panel)} and 
$1-\Gamma/\Gamma_{\rm p}$ \emph{(right panel)}
as a function of the ratio $\Gamma/M$
%normalized top right coupling $g_{t_R}$
for two values of the renormalized on-shell mass,
$M=1\ \textrm{TeV}$ (solid red lines) and $5\ \textrm{TeV}$
(dashed blue lines).}
\label{fig:respar}
\end{figure}

In order to see the differences between the two approaches,
let us numerically compare the values of the renormalized on-shell mass $M$
%(\emph{i.e.}~renormalized mass in the \emph{on-shell} approach)
and width $\Gamma$ with those of the pole mass $M_{\rm p}$
and width $\Gamma_{\rm p}$ (at the subtraction point $q_0=M$)
as a function of the right-handed top normalized coupling constant $g_{t_R}$
for different values of $M$.
As an example,
for the benchmark point $(M,g_{t_R})=(2\ \textrm{TeV},5)$,
in the on-shell approach
one gets, based on Eq.~(\ref{BWwidth})~\footnote{
The value of the QCD coupling constant is taken to be
$g_s(M_Z^2)=1.2172\pm 0.0052$, or $\alpha_s(M_Z^2)=0.1179\pm 0.0010$,
which leads after the renormalization group equation (RGE) running to
$g_s(M^2)=1.027(0.985)$ for $M=2(5)\ \textrm{TeV}$.
Hereafter, for simplicity, we consider $g_s(M^2)\simeq 1$
in all numerical calculations.},
\begin{equation}
\label{BWwidthLO}
\begin{array}{rcl}
\Gamma
&=&\sum_q\Gamma(G\to q\bar q)
=\dfrac{g_s^2}{24\pi}M\sum_q
\sqrt{1-\dfrac{4m_q^2}{M^2}}\\[4ex]
&&\times\left[g_{q_V}^2\left(1+\dfrac{2m_q^2}{M^2}\right)
+g_{q_A}^2\left(1-\dfrac{4m_q^2}{M^2}\right)\right]
=0.371\ \textrm{TeV}\ .
%\\[4ex]
%&\simeq&\dfrac{g_s^2\,g_{t_R}^2}{48\pi}
%M_{\rm BW}\left(1-\dfrac{m_t^2}{M_{\rm BW}^2}\right)
%\sqrt{1-\dfrac{4m_t^2}{M_{\rm BW}^2}}
%=0.324\ \textrm{TeV}\ .
\end{array}
\end{equation}
In the pole approach, using Eq.~(\ref{propIIRSsp}) for the determination
of the pole parameters in the second Riemann sheet of the complex plane,
one gets $(M_{\rm p},\Gamma_{\rm p})\simeq (1.96,0.387)\ \textrm{TeV}$
for the pole mass and width, respectively.
If, instead of the relativistic definition
$s_{\rm p}\equiv M_{\rm p}^2-i M_{\rm p}\Gamma_{\rm p}$,
the non-relativistic inspired one
$s_{\rm p}=(M_{\rm p}-i\Gamma_{\rm p}/2)^2$ is used,
the result is modified by less than five per mille.
This modification is stable under changes in $M$
but grows as $g_{t_R}$ increases.
For the highest values of the right-handed top coupling considered,
$g_{t_R}\simeq 10$, the modification raises up to 6\%.
In Fig.~\ref{fig:respar}, we plot 
the relative differences between the pole and renormalized on-shell
masses \emph{(left panel)}, $1-M/M_{\rm p}$,
and corresponding widths \emph{(right panel)}, $1-\Gamma/\Gamma_{\rm p}$,
for the two values $M=1\ \textrm{TeV}$ and $5\ \textrm{TeV}$, 
as a function of the ratio $r\equiv\Gamma/M$ in the range
$r\in [0.1,0.8]$~\footnote{The corresponding
right-handed top coupling $g_{t_R}$ can be obtained from Eq.~(\ref{BWwidthLO}),
once the renormalized on-shell mass $M$, the ratio $r$
and the rest of chirality quark couplings are fixed.}.
%top right coupling in the range $g_{t_R}\in [0,10]$.
%
As it can be seen, the renormalized on-shell mass 
is always bigger than the pole mass,
$M>M_{\rm p}$, for any value of the ratio $r$, with 
the absolute maximum relative difference being around 45\%(41\%)
at $r\simeq 0.8$ for $M=1(5)\ \textrm{TeV}$, respectively.
Regarding the widths,
the behavior depends upon the value of $r$: 
for $r\lesssim 0.45$, the widths satisfy $\Gamma<\Gamma_{\rm p}$,
reaching a difference of around 5\%(4\%) at $r\simeq 0.25$,
%for $M=1(5)\ \textrm{TeV}$, respectively,
while for $r\gtrsim 0.45$ one finds $\Gamma>\Gamma_{\rm p}$,
reaching an absolute maximum difference of around 36\%(26\%)
at $r\simeq 0.8$.

%\textcolor{red}{------------Aqui ---------------}

\section{Top-quark pair production at the LHC}
\label{sec:top_pair}
In the previous section, the on-shell and pole approaches
have been compared in the construction of the heavy KK-gluon propagator.
In this section, we consider the different effects of the two approaches
%(the effective one is also included for comparison)
in a measurable physical process, the top-quark pair production at the LHC,
that is, the $pp\to t\bar t X$ reaction,
mediated by the SM and the KK gluon $G$,
where $X$ represents unobserved hadrons.

The hadronic cross-section can be written schematically as
\cite{Eichten:1984eu}
\be
\sigma(pp\to t\bar t X)=\int_0^s d\hat s
\int_{\frac{1}{2}\log(\hat s/s)}^{\frac{1}{2}\log(s/\hat s)}\frac{dy}{s}
\sum_{a,b=q,\bar q,g}f^{(p)}_a(x_1) f^{(p)}_b(x_2)
\hat\sigma(ab\to t\bar t)\ ,
\label{eq:diffCS}
\ee
where $\hat\sigma(ab\to t\bar t)$ is the partonic cross-section and 
$f^{(p)}_a(x_1)dx_1$ is the probability that a parton of type $a$
carries a fraction of the incident proton momentum that lies between
$x_1$ and $x_1+dx_1$ (similarly for partons within the other incident proton).
Moreover,
$\hat s$ is the squared invariant mass of the $t\bar t$ system, 
$\hat s=x_1 x_2 s$, where $s=(13\ \textrm{TeV})^2$ is
the squared center-of-mass energy of the proton-proton system, 
$x_{1,2}=\sqrt{\hat s/s}\,e^{\pm y}$, and
$y$ is the rapidity of the top-quark pair.

To illustrate the effects described in the previous section,
we first discuss  the partonic cross-section $q\bar q\to G^\ast\to t\bar t$
for $q\neq t$~\footnote{The
case $a,b=g$ does not contribute to the KK-gluon cross-section
since the vertex $gg G$ vanishes.}.
%\footnote{The case $a,b=g$ only contributes to the SM cross-section
%since the vertex $ggG$ vanishes.}

\subsection{The partonic cross-section}
The cross-section of the Drell-Yan process at the partonic level
$q\bar q\to G^\ast\to t\bar t$ is given by
\begin{equation}
\label{eq:partonicCS}
\begin{array}{rcl}
\hat\sigma(q\bar q\to G^\ast\to t\bar t)&=&
\dfrac{g_s^4(M)}{54\pi}\dfrac{\hat{s}}{|D(\hat{s})|^{2}}
\sqrt{1-\dfrac{4m_{t}^{2}}{\hat{s}}}\\[4ex]
&&\times (g_{q_V}^2+g_{q_A}^2)
\left[g_{t_V}^2\left(1+\dfrac{2m_t^2}{\hat{s}}\right)
+g_{t_A}^2\left(1-\dfrac{4m_t^2}{\hat{s}}\right)\right]\ ,
\end{array}
\end{equation}
where two different versions of the KK-gluon propagator denominator, $D(\hat s)$, are to be assessed
\be
\left\{
\begin{array}{ll}
\textrm{\small Full propagator:} & 
\quad
D_{\rm full}(\hat s)=\hat s-M^2 -\hat s\widehat\Pi(\hat s)\ ,\\[2ex]
\textrm{\small BW propagator:} & 
\quad
D_{\rm os}(\hat s)=\hat s-M^2+i M\Gamma\ .
\end{array}
\right.
\label{propagators}
\ee
\begin{figure}
    \centering
    \includegraphics[width=7.5cm]{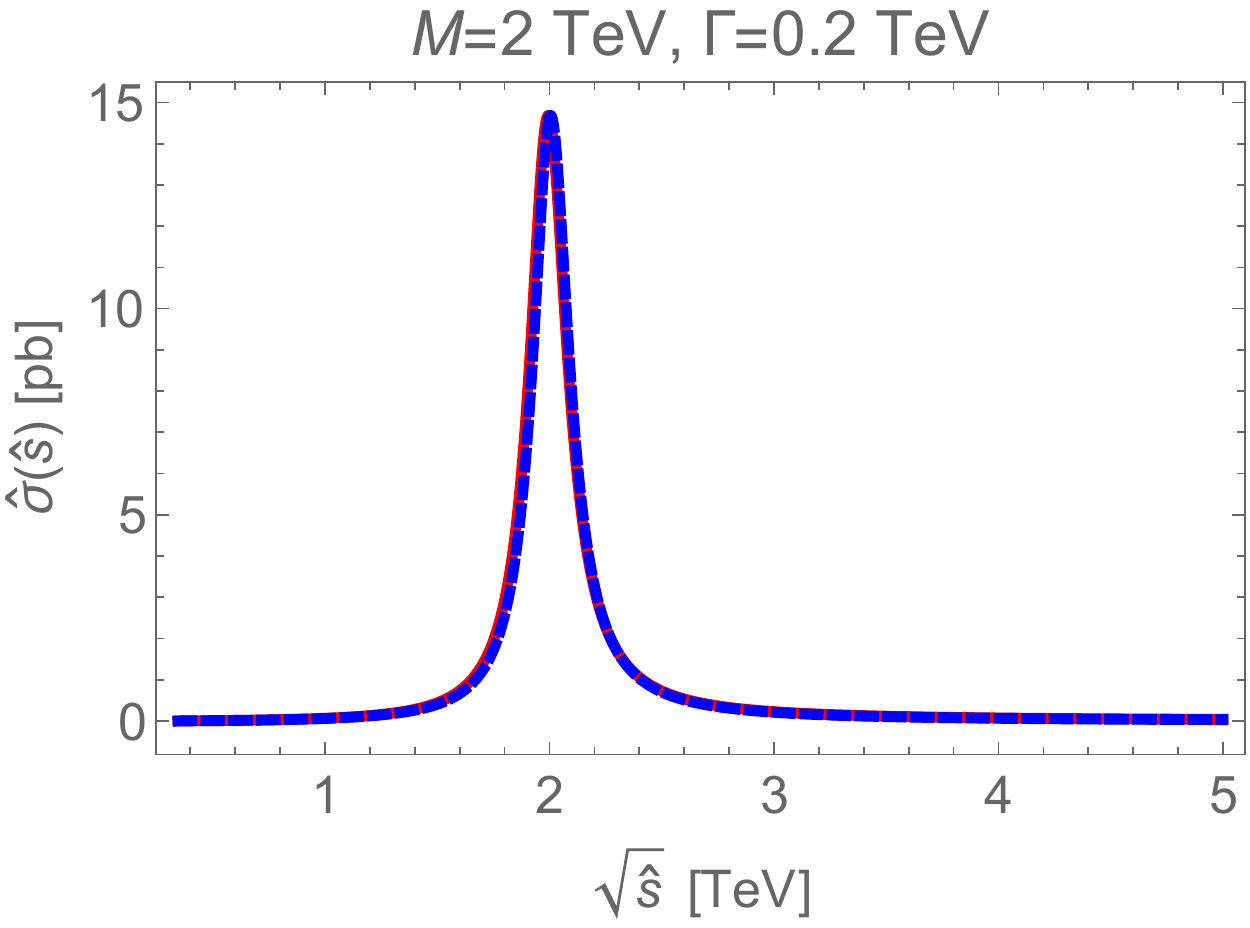}\quad
    \includegraphics[width=7.5cm]{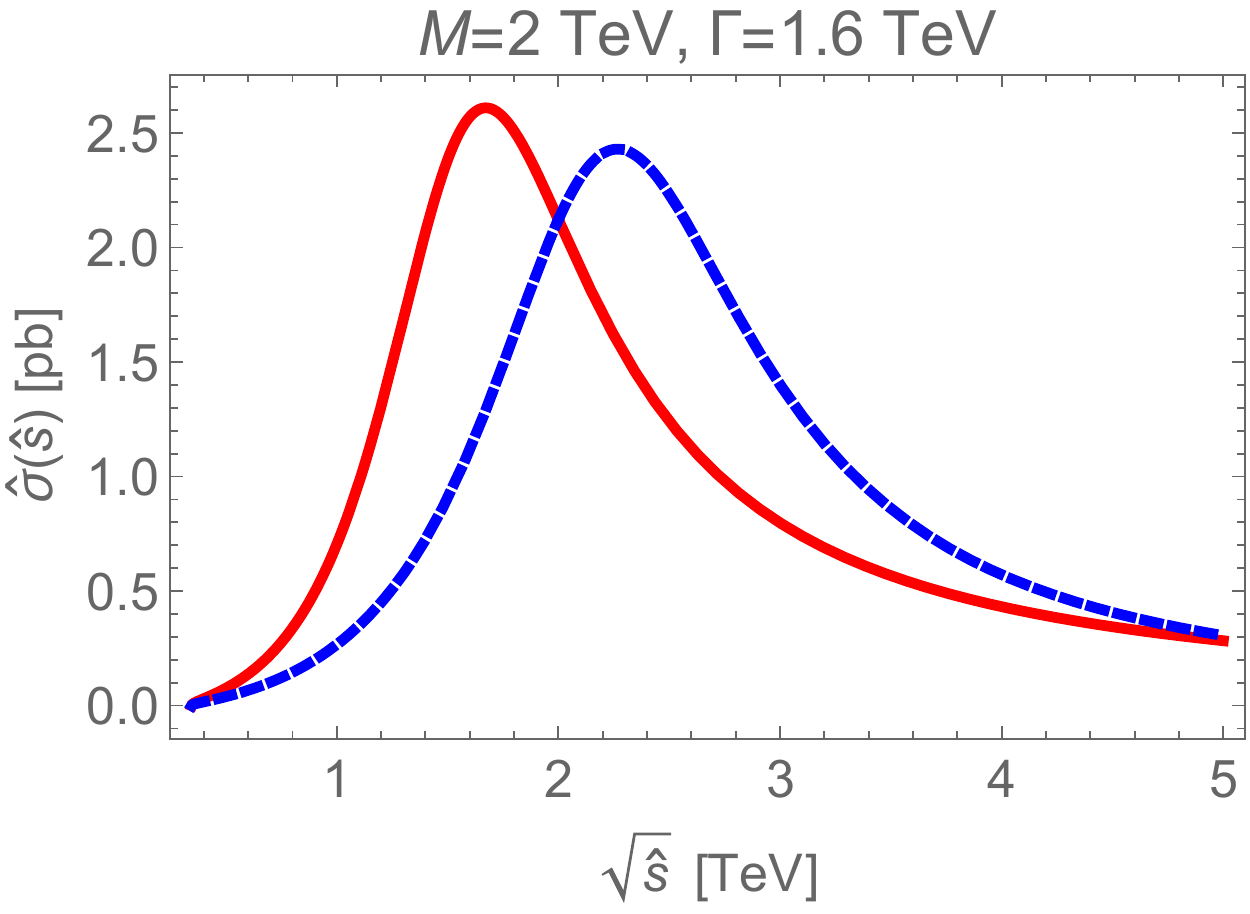}
    \caption{\it $q\bar q\to G^\ast\to t\bar t$ partonic cross-section
    as a function of the invariant mass of the top-quark pair
    for a KK gluon of $M=2$ TeV
    with a ``narrow'' width of $\Gamma=200$ GeV \emph{(left panel)}
    and a ``broad'' width of $\Gamma=1.6$ TeV \emph{(right panel)},
    respectively.
    The red solid lines correspond to the full propagator and
    the blue dashed lines to the BW propagator with on-shell parameters.
    %, and the green dotted lines to the BW propagator with pole parameters
    %obtained as explained in the text.
    }
    \label{fig:cross-sections}
\end{figure}

In Fig.~\ref{fig:cross-sections}, we present, as an example,
the partonic cross-sections for the case of a KK gluon
with a renormalized on-shell mass of $M=2$ TeV
and two different widths $\Gamma$. Specifically,
on the \emph{left panel},
a relatively narrow heavy gluon with $\Gamma=200$ GeV,
i.e.~$r=0.1$, is shown and,
on the \emph{right panel}, a broad one with $\Gamma=1.6$ TeV,
i.e.~$r=0.8$, is displayed.
In both instances, we employ the two versions of the propagator (cf.~Eq.~(\ref{propagators})), 
i.e.~the full propagator $D_{\rm full}$ (red solid lines) and
the BW propagator with on-shell parameters $D_{\rm os}$ (blue dashed lines).
%and the BW propagator with pole parameters $D_{\rm pole}$
%(green dotted lines).
%In the last case, the pole parameters used are the ones corresponding to the
%renormalized mass and width chosen as examples.
As one can clearly see, 
the two cross-sections are indistinguishable for narrow resonances,
while, for broad ones,
the location and height of the peaks are different,
as well as their effective width.
This is explained as follows.
In the narrow case,
the pole parameters associated to $(M,\Gamma)=(2,0.2)$ TeV
are $(M_{\rm p},\Gamma_{\rm p})=(1.989,0.206)$ TeV,
thus making no substantial difference.
In the broad case, however, $(M,\Gamma)=(2,1.6)$ TeV 
and the corresponding pole parameters are found to be
$(M_{\rm p},\Gamma_{\rm p})=(1.405,1.256)$ TeV; 
thus, $|D_{\rm full}(\hat s)|^{-1}$ 
peaks roughly at the pole mass,
with a larger height and narrower shape than 
$|D_{\rm os}(\hat s)|^{-1}$ (which, in turn, is located at the on-shell mass),
as a result of the pole width being smaller compared to
its on-shell counterpart.
In the limit $m_t^2/\hat s\to 0$, which is a very good approximation,
the partonic cross-section is proportional to $\hat s/|D(\hat s)|^2$;
as a consequence, the two peaks are displaced to higher values of the
top-antitop invariant mass, keeping their relative 
distance approximately constant, and their corresponding heights 
are made more alike.

As explained in Sec.~\ref{poleapp},
the resonance pole position for a specific parametrization
of the propagator is obtained from $D_{\rm{II}}(s_{\rm p})=0$, with
$s_{\rm p}=M_{\rm p}^2-i M_{\rm p}\Gamma_{\rm p}$.
%(in the second Riemann sheet for the full propagator case).
For the on-shell approach,
the pole parameters are trivially identified as
$(M_{\rm p},\Gamma_{\rm p})|_{\rm os}=(M,\Gamma)$,
while for the pole approach they are extracted from 
Eq.~(\ref{propIIRSsp}). %(second Riemann sheet).
Using, again, the examples of narrow and broad resonances from above,
the pole positions in each approach are, respectively,
$(M_{\rm p},\Gamma_{\rm p})|_{\rm os}=(2,0.2)\ \textrm{TeV}$ and
$(M_{\rm p},\Gamma_{\rm p})|_{\rm pole}=(1.989,0.206)\ \textrm{TeV}$
for $(M,\Gamma)=(2,0.2)\ \textrm{TeV}$,
and
$(M_{\rm p},\Gamma_{\rm p})|_{\rm os}=(2,1.6)\ \textrm{TeV}$ and
$(M_{\rm p},\Gamma_{\rm p})|_{\rm pole}=(1.405,1.256)\ \textrm{TeV}$
for $(M,\Gamma)=(2,1.6)\ \textrm{TeV}$.
Since the pole position in the complex plane identifies a resonance
with its mass and width,
we see that, for the same renormalized $(M,\Gamma)$ values,
one can obtain two different sets of pole parameters,
and, therefore, allegedly two different KK gluons
depending on the propagator parametrization employed.
This, clearly, is nonsense: the pole position of the KK gluon is unique
and, thus, independent of the parametrization used.
Similarly, if one instead fixes $(M_{\rm p},\Gamma_{\rm p})$, then
two different sets of KK-gluon cross-section results can be obtained
depending on the chosen propagator parametrization that 
would then need to be confronted with the available experimental data.
Given this situation, we strongly recommend the use of the pole approach,
that is, the full propagator, as it is the most complete and 
appropriate parametrization of the resonance.
The on-shell approach,
based on a BW description of the resonance,
which is considered the simplest and, therefore, easiest to implement,
is only a valid approximation for narrow
resonances which are far from other resonances and thresholds; 
accordingly, we strongly discourage using it, specially in the 
context of the present work.

\subsection{The proton cross-section}
\label{subsec:ppCS}
In this subsection, the cross-section for top-quark pair production
in proton-proton collisions via an off-shell KK gluon produced
in the $s$-channel, $pp\to G^\ast\to t\bar t$,
is computed for several values of the renormalized mass $M$ and the ratio
$r$, Eq.~(\ref{eq:ratio}), using  {\sc MadGraph5$_-$aMC}~\cite{Alwall:2014hca}.
We provide predictions for the two different propagator
parametrizations discussed above, that is,
the BW propagator with on-shell parameters and the full propagator,
and compare our results with experimental data when available.

\begin{figure}[htb]
    \centering
    \includegraphics[width=7.5cm]{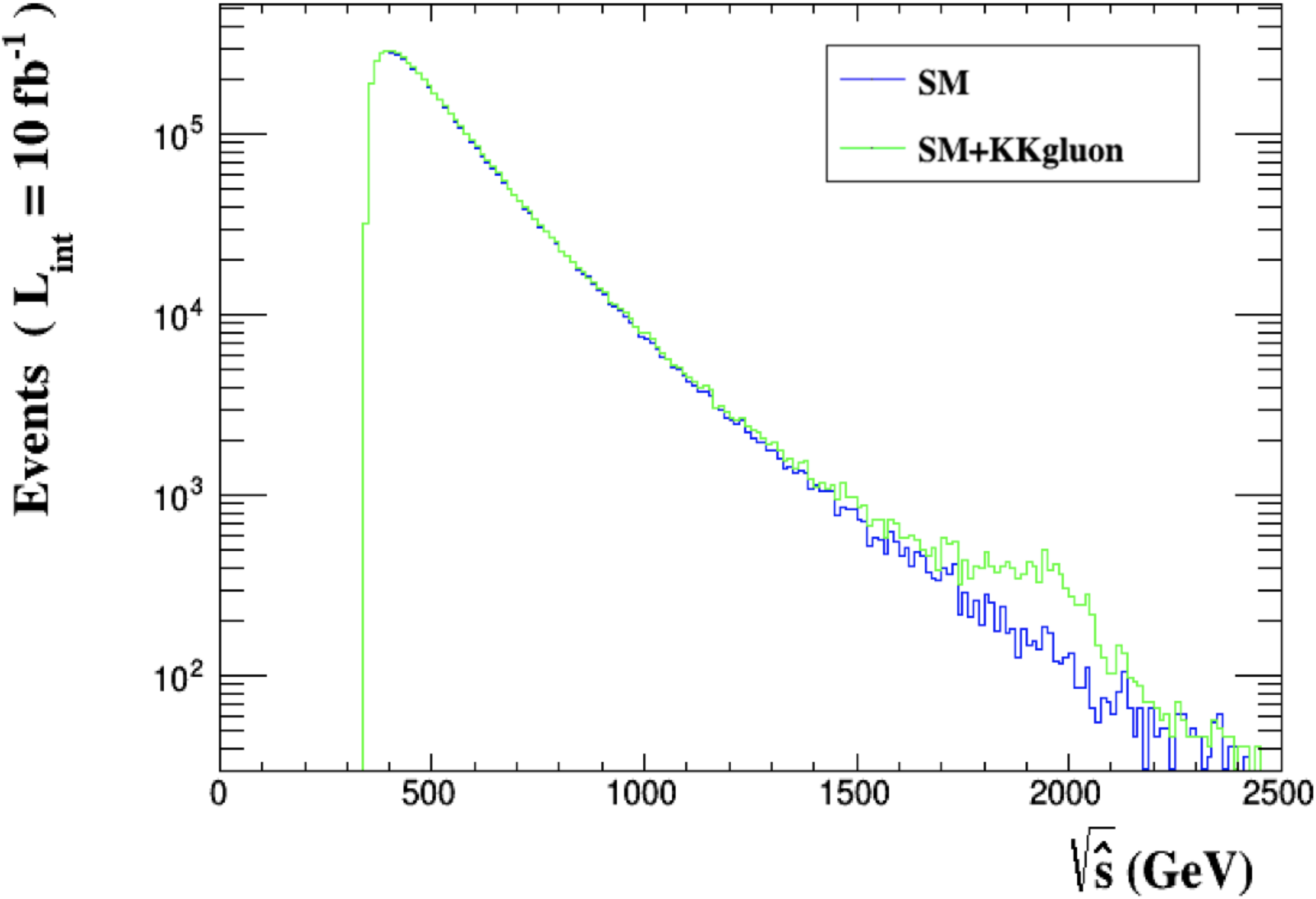}\quad
    \includegraphics[width=7.5cm]{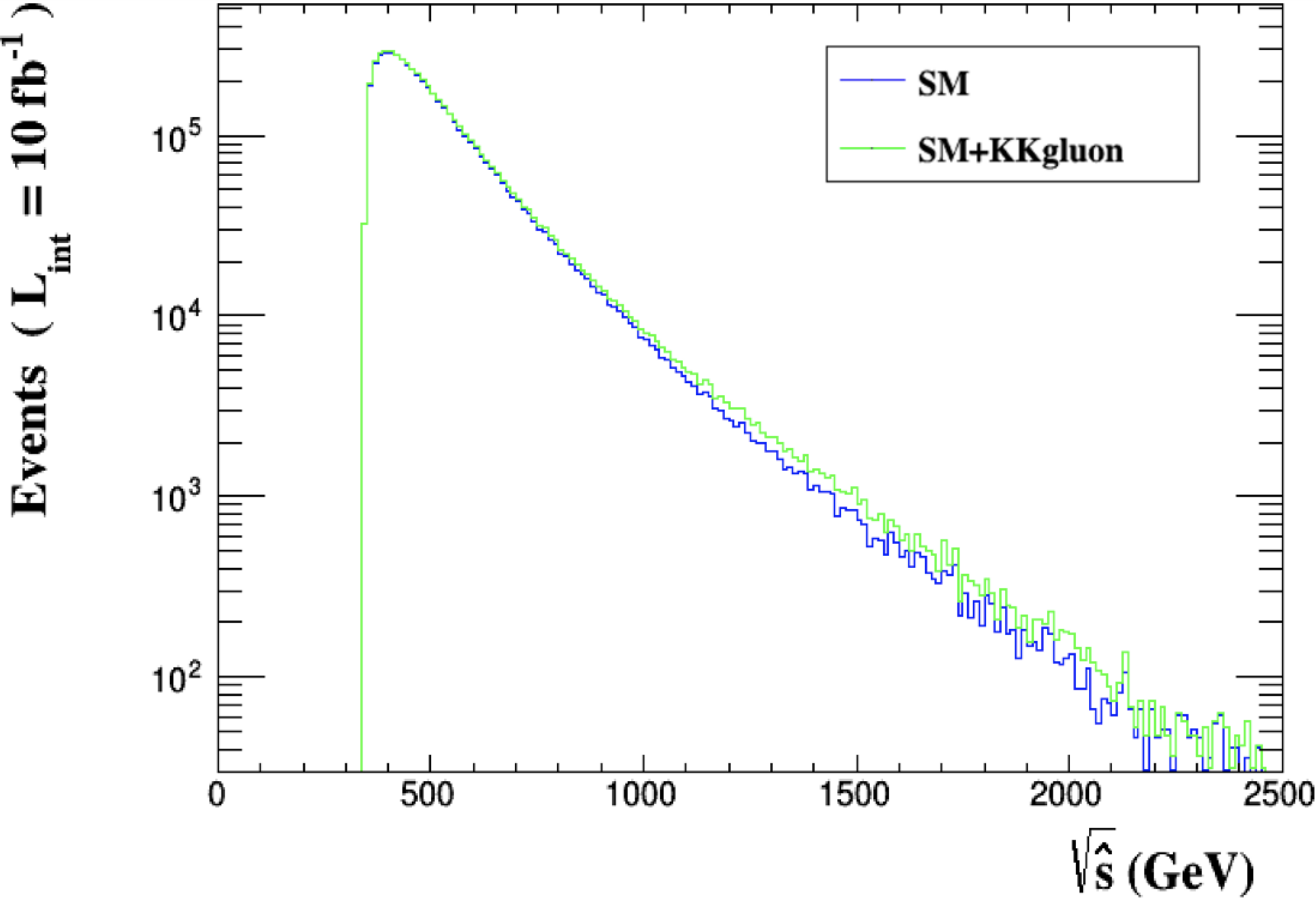}
    \caption{\it $pp\to t\bar t$ integrated cross-section
    as a function of the top-antitop invariant mass $\sqrt{\hat s}$
    for $r=0.1$ \emph{(left panel)} and $r=0.8$ \emph{(right panel)}.
    In both cases, the renormalized on-shell mass is set to
    $M=2\ \textrm{TeV}$.
    The number of distributed events is $10^6$ and
    the integrated luminosity is assumed to be $10\ \textrm{fb}^{-1}$.
    The blue lines correspond to the SM contribution alone and
    the green lines to the SM and KK-gluon contributions.}
    \label{fig:ppttCS3}
\end{figure}

First of all, and in order to illustrate the effect of broad resonances on experimental detection, we display in Fig.~\ref{fig:ppttCS3}
the proton-proton integrated cross-section as a function of 
the top-antitop invariant mass $\sqrt{\hat s}$
for $r=0.1$ \emph{(left panel)} and $r=0.8$ \emph{(right panel)}.
In both cases, the renormalized on-shell mass is set to
$M=2\ \textrm{TeV}$.
The number of distributed events is $10^6$ and
the integrated luminosity is assumed to be $10\ \textrm{fb}^{-1}$.
The calculation performed by 
{\sc MadGraph5$_-$aMC}~\cite{Alwall:2014hca} of both
the SM background and the KK-gluon signal is at leading order.
%and taking only into account the $s$-channel contribution of the
%QCD gluon \textcolor{red}{in Drell-Yan processes}.
The blue lines correspond to the SM contribution alone and
the green lines to the SM and KK-gluon contributions, where the
KK-gluon contributions are calculated using the full propagator.
As shown, for $r=0.1$, or equivalently $\Gamma=200\ \textrm{GeV}$,
the effect of the narrow KK gluon is clearly seen as a bump
at $\sqrt{\hat s}\simeq 2\ \textrm{TeV}$ standing up over the 
SM background.
This behavior makes it easier to isolate the signal and
allows for the discovery of such a narrow resonance.
On the contrary, for $r=0.8$, corresponding to a broad KK gluon with
$\Gamma=1.6\ \textrm{TeV}$, the bump disappears and its effect is
spread in a much wider region of $\sqrt{\hat s}$,
making it more difficult to extract a possible signal from the 
background.
Once again, we want to stress that Fig.~\ref{fig:ppttCS3} is for illustrative purpose only.

\begin{figure}[htb]
    \centering
    \includegraphics[width=7.5cm]{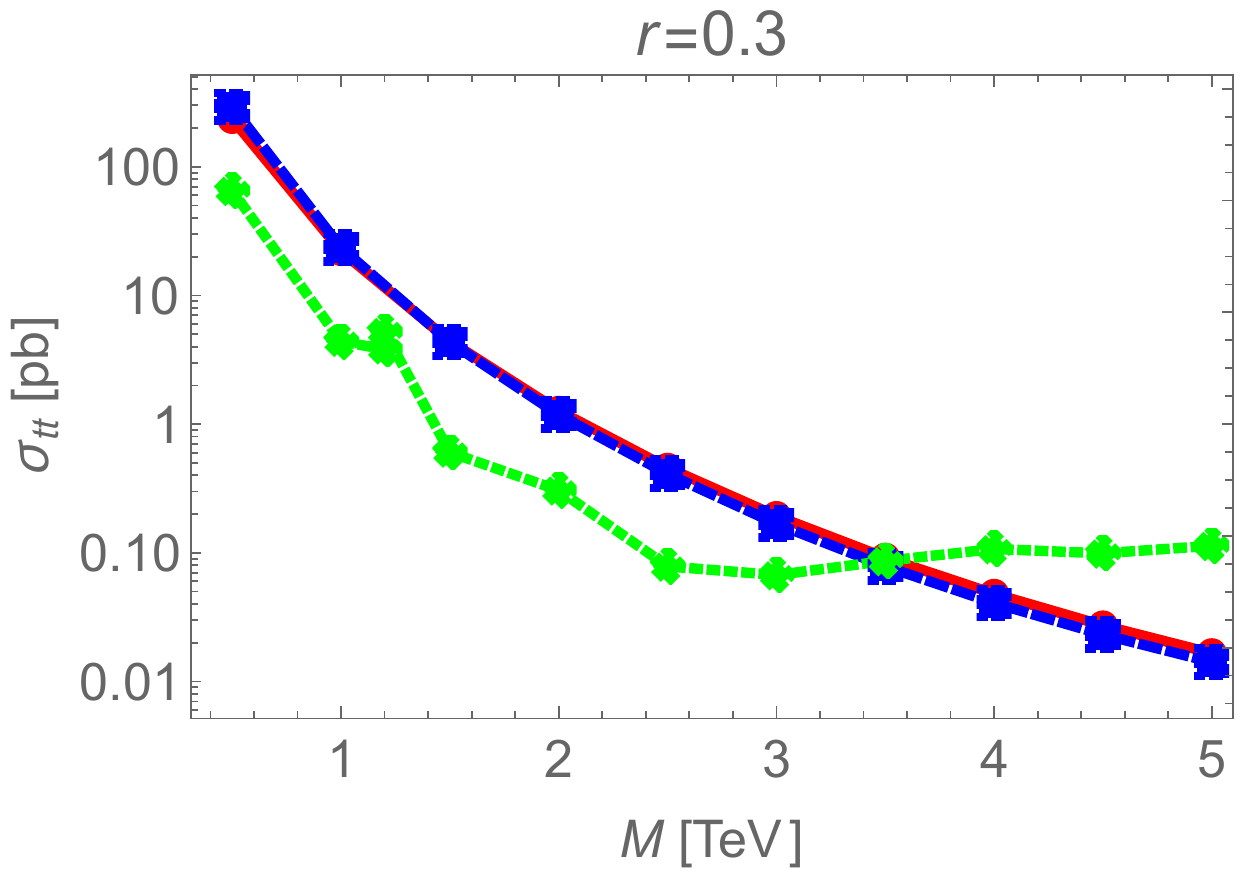}\quad
    \includegraphics[width=7.5cm]{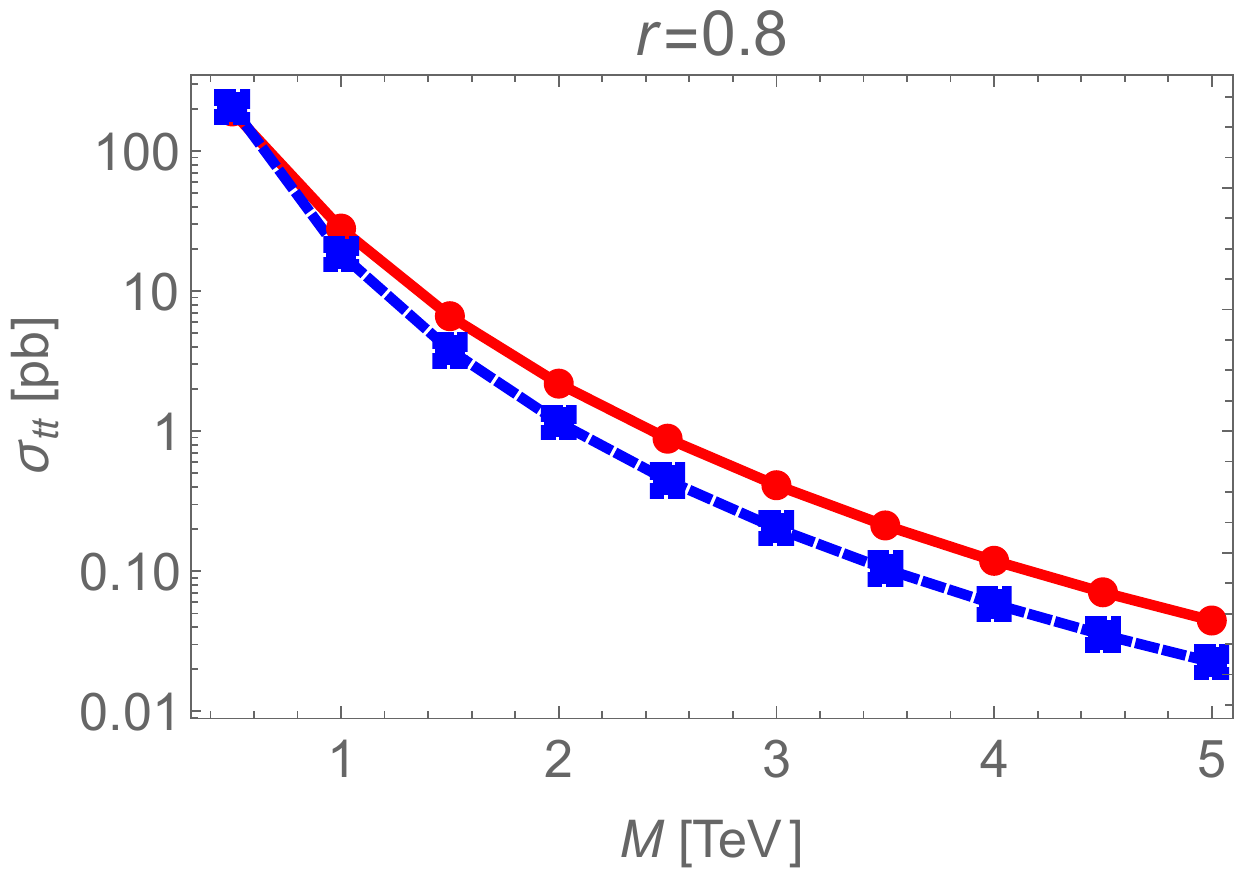}
    \caption{\it $pp\to G^\ast\to t\bar t$ integrated cross-section
    as a function of the renormalized on-shell mass $M$ for the ratio
    $r=0.3$ \emph{(left panel)} and $r=0.8$ \emph{(right panel)}.
    The red solid lines correspond to the full propagator,
    the blue dashed lines to the BW propagator with on-shell parameters, 
    and the green dotted line to the experimental 95\% CL upper limit
    observed by the ATLAS collaboration~\cite{Aaboud:2019roo}.}
    \label{fig:ppttCS}
\end{figure}

In Fig.~\ref{fig:ppttCS},
we display the proton-proton integrated cross-section as a function of $M$
for $r=0.3$ \emph{(left panel)} and $r=0.8$ \emph{(right panel)}.
The red solid lines correspond to the predictions based on the full propagator, 
while the blue dashed ones are based on the BW propagator with on-shell parameters.
The cross-sections are calculated for the same values as the ATLAS collaboration in 
Ref.~\cite{Aaboud:2019roo}, namely,
from $M=0.5\ \textrm{TeV}$ to $5\ \textrm{TeV}$ in steps of $0.5\ \textrm{TeV}$.
All the values shown are computed from simulations with $10^5$ events.
As expected, the predicted values decrease when 
the KK-gluon on-shell mass increases.

For the $r=0.3$ case,
the predictions for the two types of propagators are almost indistinguishable,
the reason being the low value of the KK-gluon on-shell width $\Gamma$
with respect to its mass $M$,
a behaviour already anticipated at the partonic level.
These two sets of predictions are compared with the experimental
95\% CL upper limits observed by the ATLAS collaboration in 
Ref.~\cite{Aaboud:2019roo} (green dotted line).
Note that the two experimental values for the 
cross-section at $M=1.2\ \textrm{TeV}$
correspond to the resolved and boosted analysis methods taken by ATLAS.
Our values using the BW propagator are very similar to the theoretical
prediction displayed in Fig.~12 of Ref.~\cite{Aaboud:2019roo} by the red line.
As it can be seen in Fig.~\ref{fig:ppttCS}, values of the KK-gluon on-shell mass 
$\lesssim 3.5\ \textrm{TeV}$ for a resonance with $r=0.3$ 
are excluded at the 95\% CL for both propagator parametrizations.

For the $r=0.8$ case, the two sets of predictions are clearly distinguishable,
with the cross-section values from the full propagator 
being always bigger than those of
the BW propagator for all values of $M$. 
There is a simple explanation for this behavior: 
as we have already discussed before, the pole mass is smaller than the renormalized one, $M_{\rm p}<M$; 
consequently, the maximum of the partonic cross-section computed using the full propagator takes place at a $\sqrt{\hat s}$ value 
that is smaller than the corresponding one for the BW propagator (cf.~Fig.~\ref{fig:cross-sections}). 
At the proton level, a resonance with mass $M_{\rm p}$ is thus more copiously produced than a resonance with a larger mass $M$, 
and, consequently, the cross-section computed with the full propagator is larger than the cross-section obtained with the BW propagator,
in agreement with both panels in Fig.~\ref{fig:ppttCS}. 
In addition, and as we saw in Fig.~\ref{fig:respar}, the larger the value of $r$, the smaller the $M_{\rm p}/M$ ratio, 
which in turn leads to a larger ratio of cross-sections, $\sigma_{t\bar t}^{\textrm{full}}/\sigma_{t\bar t}^{\textrm{BW}}$. 
This explains the difference between both panels in Fig.~\ref{fig:ppttCS}.
Unfortunately, there currently is no
experimental data for $r=0.8$ to compare with our results.
Notwithstanding this, one can see that an experimental exclusion line
would exclude smaller values of $M$ in the case of the BW propagator
as compared to the full propagator. In fact, if one assumes an approximately 
flat crossing, then the difference would be of order $0.5\ \textrm{TeV}$
for $M\ge 1.5\ \textrm{TeV}$.

\begin{figure}[htb]
    \centering
    \includegraphics[width=7.5cm]{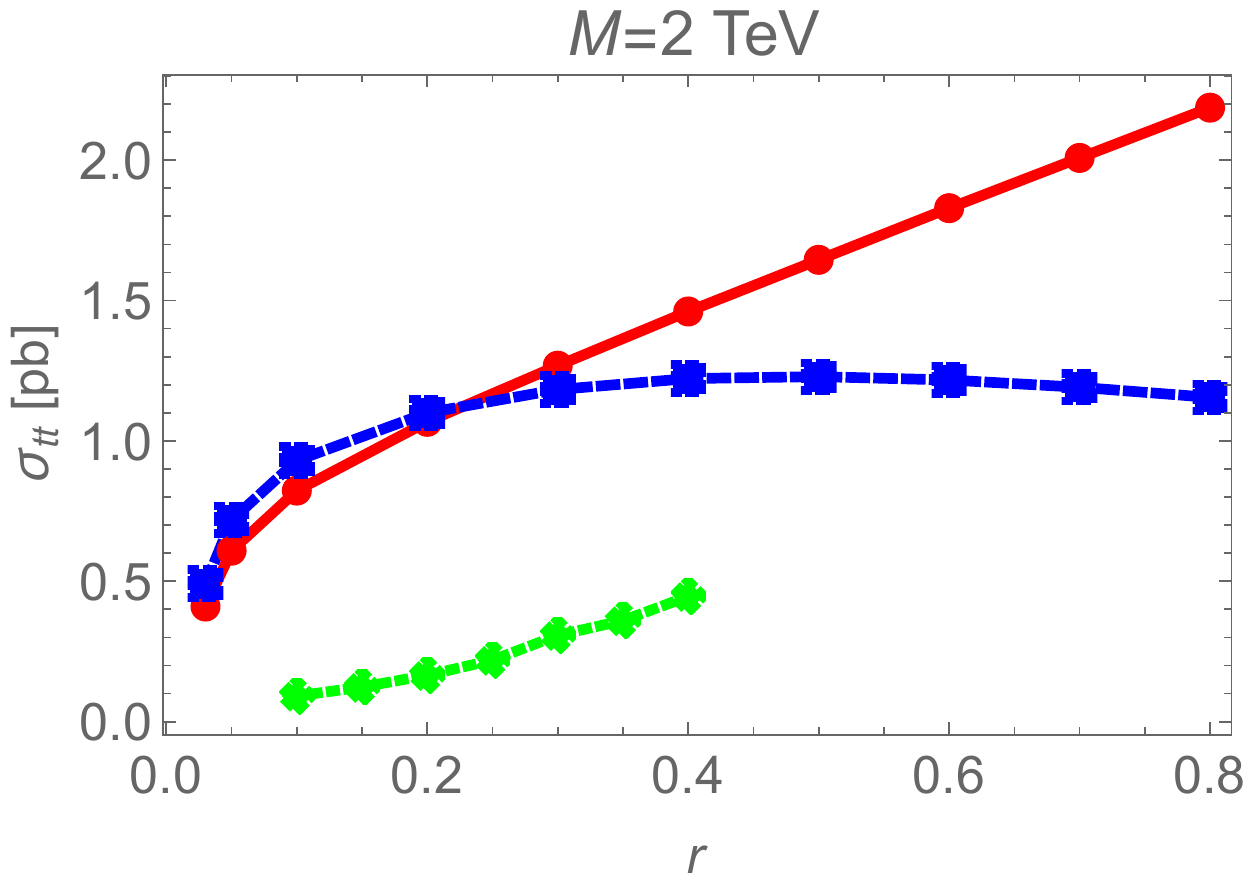}\quad
    \includegraphics[width=7.5cm]{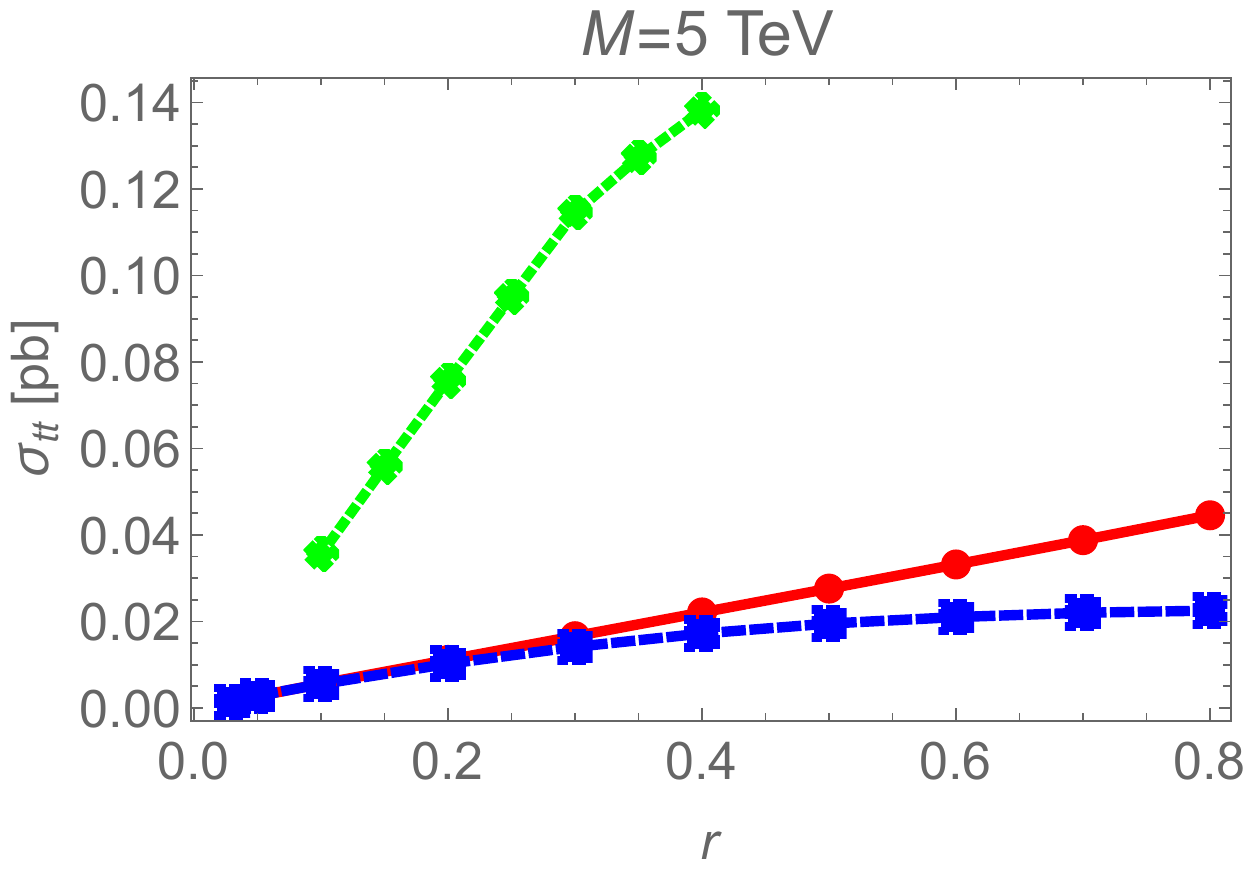}
    \caption{\it $pp\to G^\ast\to t\bar t$ integrated cross-section
    as a function of the ratio $r=\Gamma/M$ for the renormalized
    on-shell mass $M=2\ \textrm{TeV}$ \emph{(left panel)} and
    $5\ \textrm{TeV}$ \emph{(right panel)}.
    The red solid lines correspond to the full propagator,
    the blue dashed lines to the BW propagator with on-shell parameters, 
    and the green dotted line to the experimental 95\% CL upper limit
    observed by the ATLAS collaboration~\cite{Aaboud:2019roo}.}
    \label{fig:ppttCS2}
\end{figure}

Similarly, in Fig.~\ref{fig:ppttCS2}
we display the proton-proton integrated cross-section as a function of $r$
for $M=2\ \textrm{TeV}$ \emph{(left panel)} and 
$5\ \textrm{TeV}$ \emph{(right panel)}.
The cross-sections are calculated for values of $r$ ranging from
$r=0.1$ to $r=0.8$, while the experimental ATLAS values~\cite{Aaboud:2019roo} range from $r=0.1$ to $r=0.4$.
We also include predicted values for $r=0.03$ and $r=0.05$
%from $r=0.01$ to $r=0.03$
to test the convergence of both approaches for very narrow resonances.
Using the full propagator, the predicted cross-sections increase as $r$, 
or equivalently $\Gamma$, increases; this is the case for 
the two values of $M$ analyzed here.
Contrary to this, when the BW propagator is used, 
the predicted cross-sections appear to approach a constant value
for $r\gtrsim 0.4$.
Qualitatively, the results from Fig.~\ref{fig:ppttCS2} can be explained by the fact that the pole mass is always smaller than the renormalized one, 
$M_{\rm p}<M$, and, as already discussed above, the larger $r$ is, the smaller the ratio $M_{\rm p}/M$ becomes (cf.~Fig.~\ref{fig:respar}). 
This means that the ratio of cross-sections computed with full and BW propagators, 
$\sigma_{t\bar t}^{\textrm{full}}/\sigma_{t\bar t}^{\textrm{BW}}$,
increases for larger values of $r$, in agreement with both panels in Fig.~\ref{fig:ppttCS2}. 
Furthermore, the left panel in Fig.~\ref{fig:respar} also shows that the ratio $M_{\rm p}/M$ decreases for increasing values of $M$, 
which helps explain the difference between the panels in Fig.~\ref{fig:ppttCS2}.
In view of the experimental exclusion lines shown in Fig.~\ref{fig:ppttCS2},
a KK gluon with an on-shell mass of $M=2\ \textrm{TeV}$ is excluded,
%no matters what the associated width is, 
at least for any $r\leq 0.4$. Conversely, a KK gluon 
with an on-shell mass of $M=5\ \textrm{TeV}$ is allowed for any width.
This is in line with the conclusions obtained from the left panel in
Fig.~\ref{fig:ppttCS} whereby a possible exclusion
starts at $M\simeq 3.5\ \textrm{TeV}$ for relatively narrow $(r=0.3$) KK gluons.

\section{Top-quark pair plus $W$, $Z$ or $H$ production at the LHC}
\label{sec:WZH}
In the following subsections, we present our predictions for the total
proton-proton cross-sections into a top-antitop quark pair plus $W$, $Z$ or $H$
production at the LHC for different values of the KK-gluon on-shell mass $M$
and the ratio $r$.
All the predicted values are computed with simulations using $10^5$ events
and presented using $\mu_{\rm th}\equiv\sigma_{\rm total}/\sigma_{\rm SM}$
defined as the ratio of the total cross-section 
with respect to the SM value.
These results are compared, process by process, with available
experimental data from the ATLAS or CMS collaborations
in the form of $\mu_{\rm obs}\equiv\sigma_{\rm obs}/\sigma_{\rm SM}$,
where $\sigma_{\rm obs}$ is the observed cross-section.
The calculated total cross-sections include the contributions of the SM,
KK gluon and their interference.

For the SM contribution, we assume the same value the
experimental collaboration with which we compare uses,
while for the KK-gluon and the interference \footnote{The 
interference is calculated by subtracting from the 
total cross-section the sum of the SM and 
KK-gluon cross-sections.} contributions 
we use the values given by {\sc MadGraph5$_-$aMC}~\cite{Alwall:2014hca}.
Given that for each process we take the value of the
SM cross-section provided by the experimental collaborations,
the most conservative criterion followed to choose the data to compare with
is the one offering a value of $\mu_{\rm obs}$ closer to unity,
that is, more compatible with the SM.
For the sake of simplicity, we only compare with
the experimental data the predictions that we compute using
the full propagator (pole approach),
which %we believe 
is the most correct way to proceed.
For the three processes that we analyze, 
we include plots of the ratio $\mu_{\rm th}$
as a function of $M$ for fixed values of $r=0.3$ and $0.8$, and 
$\mu_{\rm th}$ as a function of $r$ for fixed $M=2\ \textrm{TeV}$ and 5 TeV.
Our predictions are shown in solid (red) and blue (dashed) lines
while the experimental data are displayed using bands corresponding to 
$1\sigma$ (light green) and $2\sigma$ (light yellow) deviations from
the $\mu_{\rm obs}$ (black dotted line) central values.
A subset of representative Feynman diagrams contributing to each process
is also shown for completeness.

\subsection{Top-quark pair plus $W$ production}
\label{ppttWCS}
\begin{figure}[htb]
    \centering
    \includegraphics[width=6.5cm]{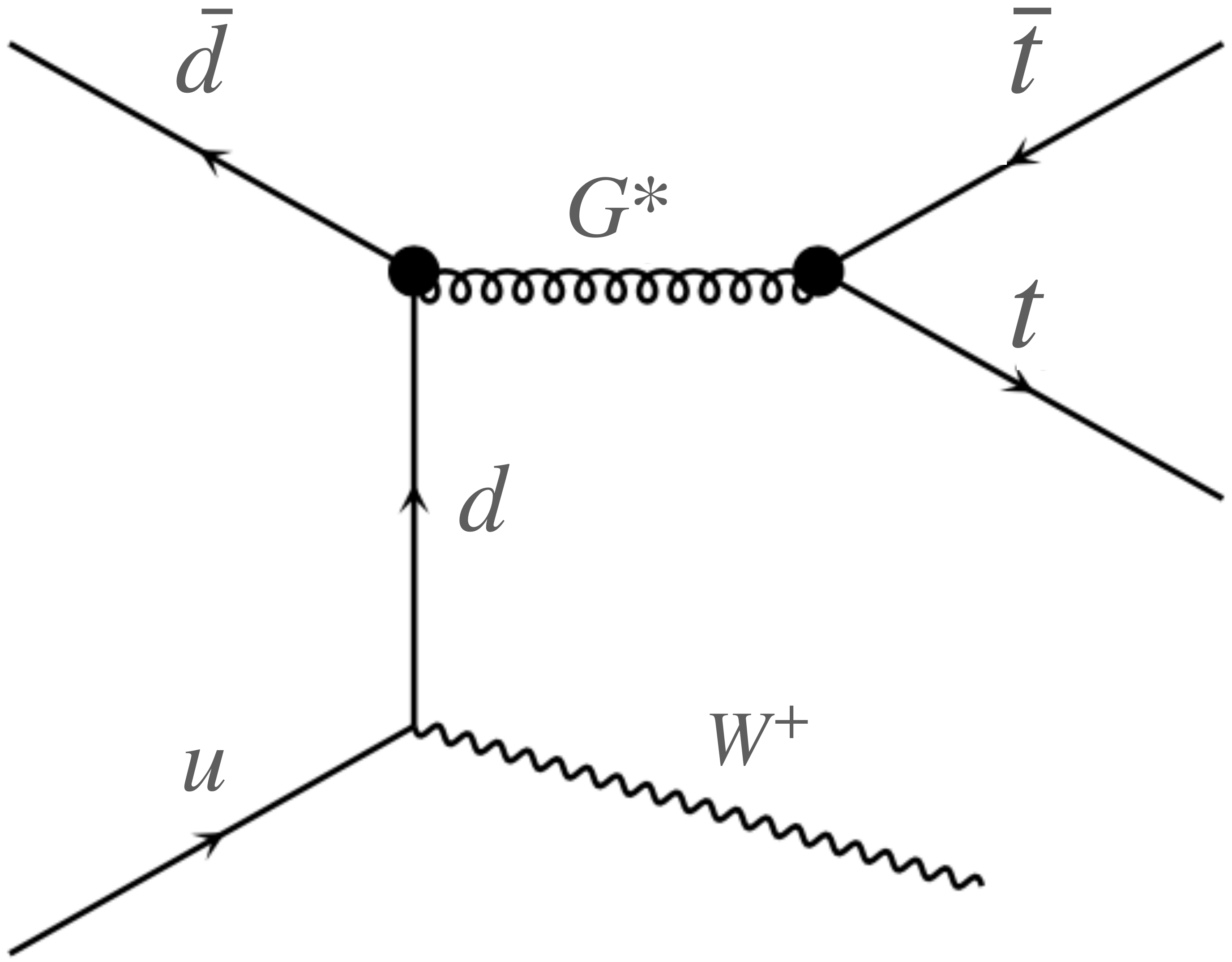}
    \caption{\it A representative Feynman diagram of the process
    $pp\to t\bar t W$ mediated by a KK gluon.
    This is only one of the 8 diagrams containing a KK gluon
    out of a total of 76 involved in the process.}
    \label{fig:ppttWCSFeynmanD}
\end{figure}

The $pp\to t\bar t W$ total cross-section involves a total of 76
Feynman diagrams, 8 of which involve a KK gluon.
An example of such diagrams is shown in Fig.~\ref{fig:ppttWCSFeynmanD}.
%The SM cross-section is calculated to be
%$\sigma_{\rm SM}^{t\bar t W}=0.234\ \textrm{pb}$ in MadGraph5.
The experimental observed ratios are
$\mu_{t\bar t W}^{\rm obs}=1.39^{+0.17}_{-0.16}$
(for a reference SM value of
$\sigma_{t\bar t W}^{\rm SM}=727\ \textrm{fb}$)
measured by the ATLAS collaboration \cite{ATLAS:2019nvo} and
$\mu_{t\bar t W}^{\rm obs}=1.43\pm 0.21$ 
(with $\sigma_{t\bar t W}^{\rm SM}=650\ \textrm{fb}$)
by the CMS collaboration~\cite{Sirunyan:2020icl}. Both measurements exhibit a mild $\sim 2\sigma$ departure from the SM prediction. 
This represents a small anomaly with respect to the SM that, if confirmed by future, more precise, data, 
should point toward the presence of new physics. 
Unexpectedly, the measurements of the observable $\mu_{t\bar t W}^{\rm obs}$ from both experimental collaborations, ATLAS and CMS, 
are consistent with each other within $1\sigma$. 
However, we prefer to conservatively wait for future confirmation or otherwise. 
%remain skeptical about it for the time being and prefer to wait for future confirmation or otherwise.

\begin{figure}
    \centering
    \includegraphics[width=7.5cm]{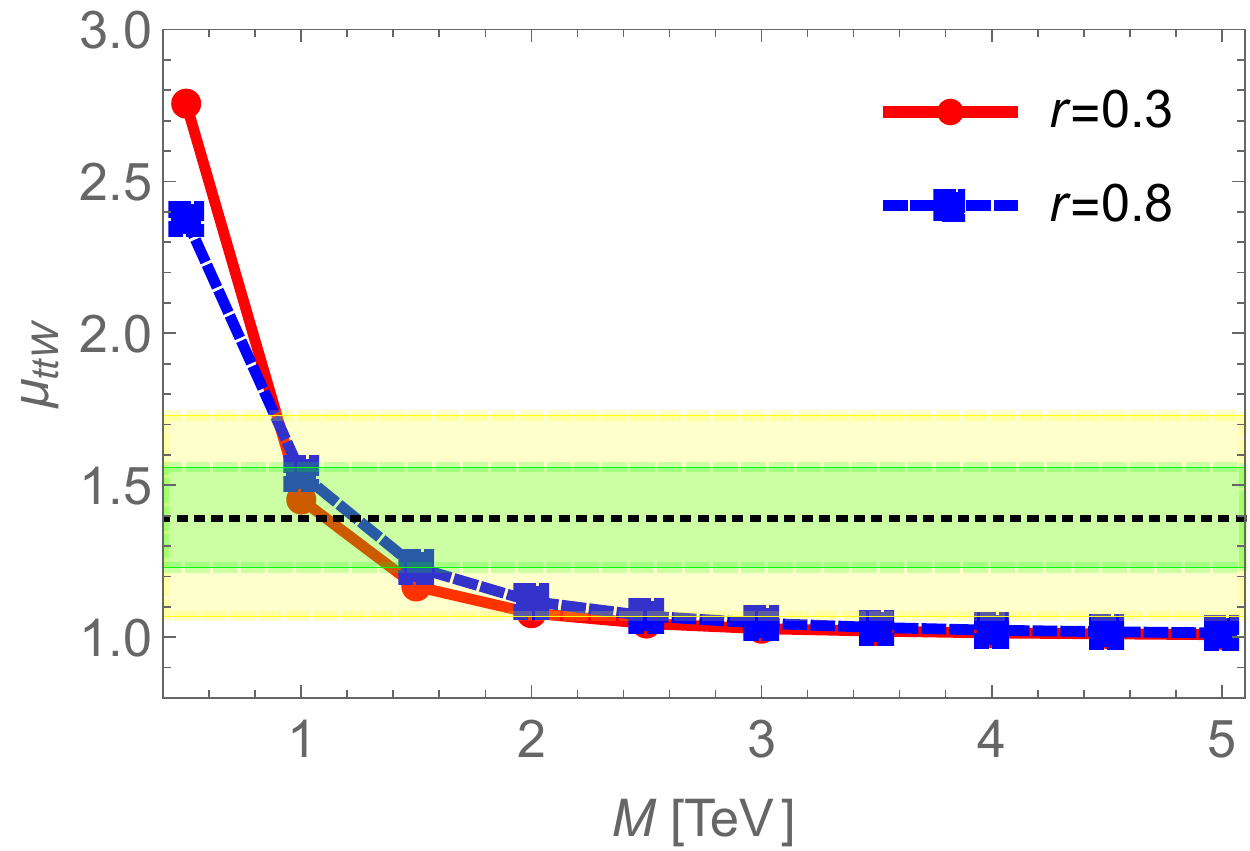}\quad
    \includegraphics[width=7.5cm]{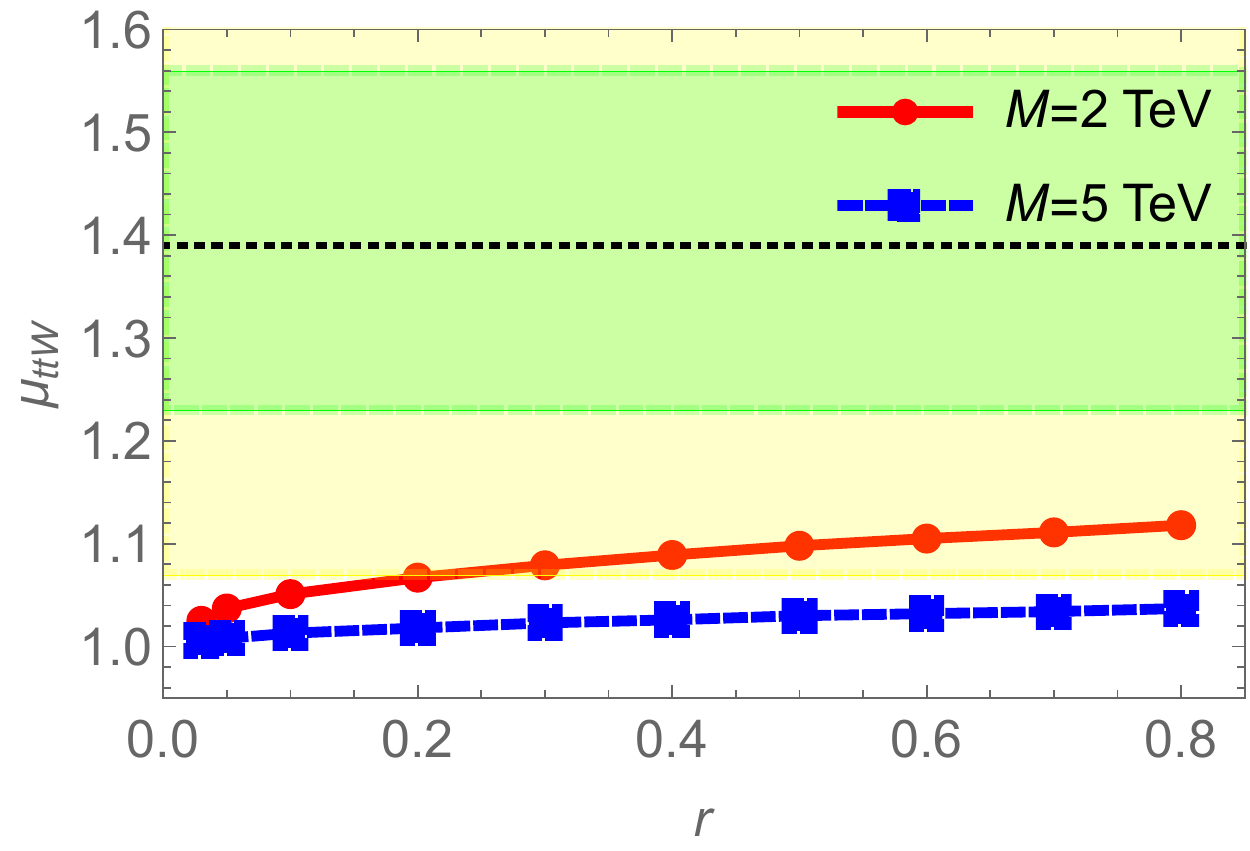}
    \caption{\it Ratio of the total cross-section $pp\to t\bar t W$
    with respect to the SM value, $\mu_{t\bar t W}^{\rm th}$,
    as a function of the renormalized on-shell mass $M$
    \emph{(left panel)} for the fixed ratios
    $r=0.3$ (red solid line) and $r=0.8$ (blue dashed line),
    and as a function of the ratio $r$ \emph{(right panel)}
    for the fixed masses $M=2\ \textrm{TeV}$ (red solid line) and
    $M=5\ \textrm{TeV}$ (blue dashed line).
    The different lines correspond to the predictions based on the
    full propagator (pole approach), while the bands represent
    $1\sigma$ (light green) and $2\sigma$ (light yellow)
    deviations from the $\mu_{t\bar t W}^{\rm obs}$ central value
    (black dotted line) observed by the ATLAS collaboration \cite{ATLAS:2019nvo}.}
    \label{fig:ppttWCS}
\end{figure}

In Fig.~\ref{fig:ppttWCS},
our predictions for $\mu_{t\bar t W}^{\rm th}$
as a function of $M$ \emph{(left panel)} and of the ratio $r$
\emph{(right panel)} are compared with the experimental value
observed by ATLAS.
Should this experimental value be confirmed, then a KK gluon of mass roughly between 1 and 1.5 TeV may be allowed at the $1\sigma$ level, regardless of its width.
However, if the bands are relaxed to include $2\sigma$ deviations,
the allowed range encompasses from about 0.8 to 2.5 TeV.
As well as this, KK-gluon masses of 2 and 5 TeV would be excluded
at $1\sigma$ irrespective of their widths,
while at $2\sigma$ only a KK gluon of $M=2\ \textrm{TeV}$ and 
$\Gamma\gtrsim 400\ \textrm{GeV}$ would be allowed. 
These results ought to be understood in conjunction and compared with the existing data on top-pair production, which yield, 
as we have seen in the previous section, the bound $M\gtrsim 3.5$ TeV for $r\leq 0.4$ at 95\% CL. 
Specifically, a KK gluon with a mass $M=5$ TeV, irrespective of $r$, should be excluded from the results presented in Fig.~\ref{fig:ppttWCS}.

\subsection{Top-quark pair plus $Z$ production}
\label{ppttZCS}

\begin{figure}[htb]
    \centering
    \includegraphics[width=12.5cm]{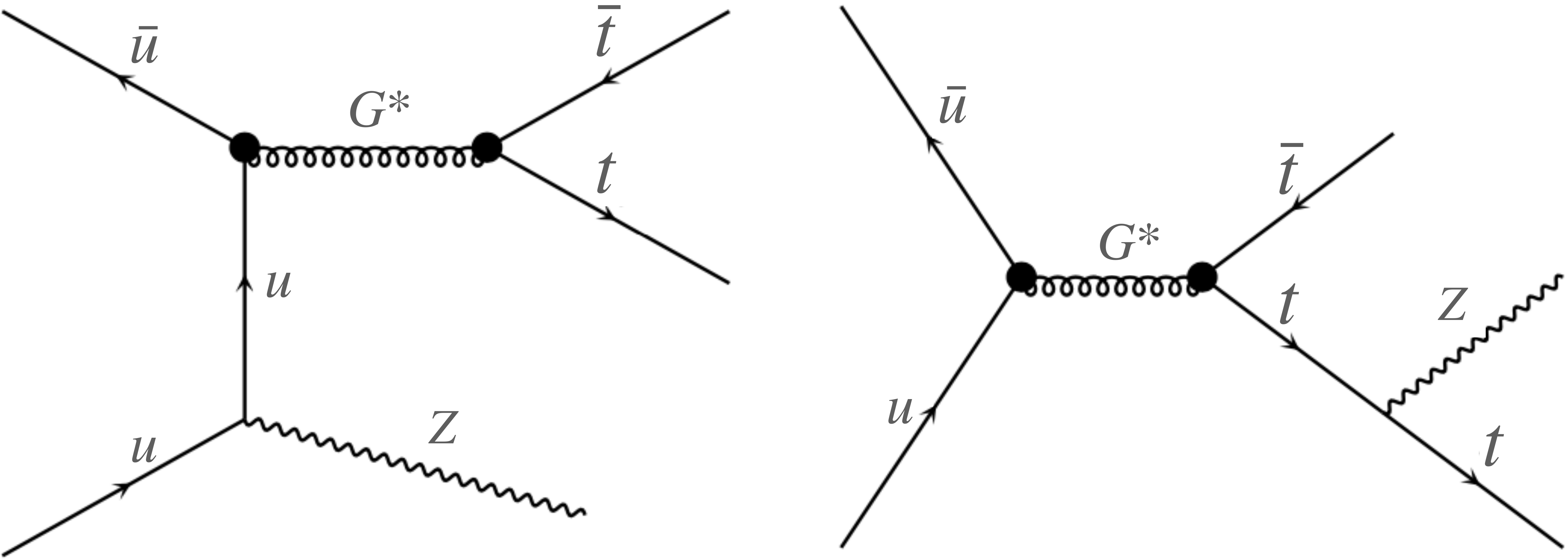}
    \caption{\it A representative subset of Feynman diagrams of the process
    $pp\to t\bar t Z$ mediated by a KK gluon.
    These are only two of the 16 diagrams containing a KK gluon
    out of a total of 116 involved in the process.}
    \label{fig:ppttZCSFeynmanD}
\end{figure}

The $pp\to t\bar t Z$ total cross-section involves a total of 116
Feynman diagrams, 16 of which involve a KK gluon.
Examples of such diagrams are shown in Fig.~\ref{fig:ppttZCSFeynmanD}.
The KK-gluon mass can be reconstructed either from the
$t\bar t$ invariant mass \emph{(left)} or the 
$t\bar t Z$ one \emph{(right)}.
%The SM cross-section is calculated to be
%$\sigma_{\rm SM}^{t\bar t Z}=0.582\ \textrm{pb}$ in MadGraph5.
The experimental observed ratios are 
$\mu_{t\bar t Z}^{\rm obs}=1.19\pm 0.12$
(for a reference SM value of
$\sigma_{t\bar t Z}^{\rm SM}=880\ \textrm{fb}$)
measured by the ATLAS collaboration \cite{ATLAS:2020cxf} and
$\mu_{t\bar t Z}^{\rm obs}=1.03\pm 0.14$
(with $\sigma_{t\bar t Z}^{\rm SM}=839\ \textrm{fb}$)
by the CMS collaboration~\cite{Sirunyan:2020icl}.
\begin{figure}[htb]
    \centering
    \includegraphics[width=7.5cm]{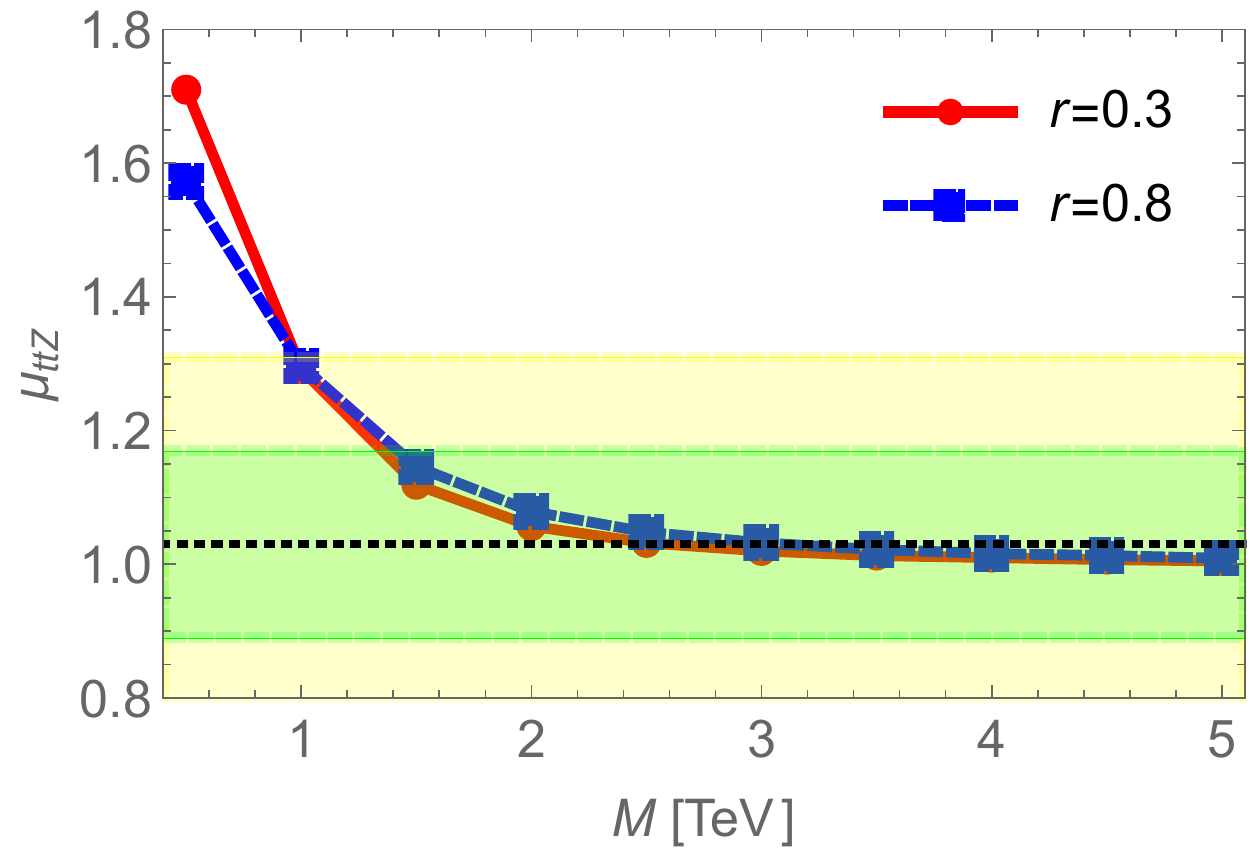}\quad
    \includegraphics[width=7.5cm]{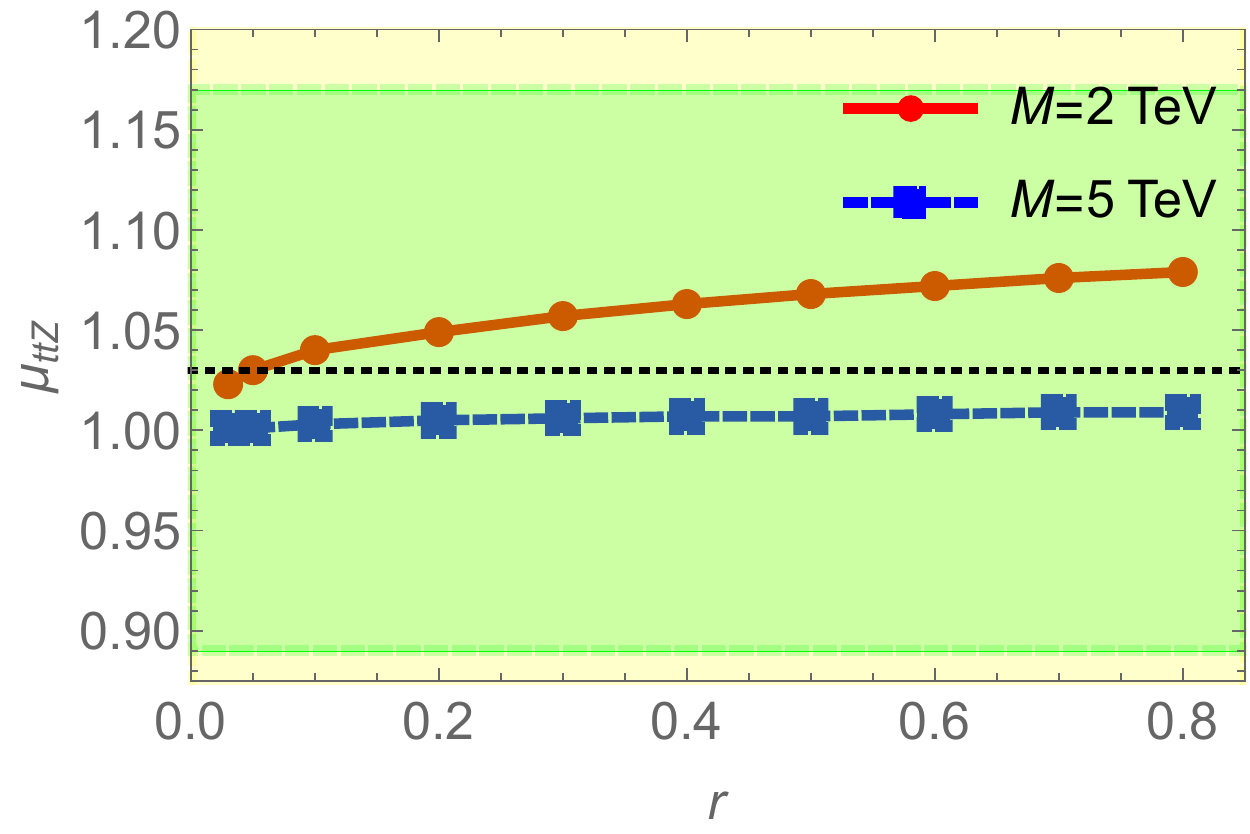}
    \caption{\it Ratio of the total cross-section $pp\to t\bar t Z$
    with respect to the SM value, $\mu_{t\bar t Z}^{\rm th}$,
    as a function of the renormalized on-shell mass $M$
    \emph{(left panel)} for the fixed ratios
    $r=0.3$ (red solid line) and $r=0.8$ (blue dashed line),
    and as a function of the ratio $r$ \emph{(right panel)}
    for the fixed masses $M=2\ \textrm{TeV}$ (red solid line) and
    $M=5\ \textrm{TeV}$ (blue dashed line).
    The different lines correspond to the predictions based on the
    full propagator (pole approach), while the bands represent
    $1\sigma$ (light green) and $2\sigma$ (light yellow)
    deviations from the $\mu_{t\bar t Z}^{\rm obs}$ central value
    (black dotted line) observed by the CMS collaboration \cite{Sirunyan:2020icl}.}
    \label{fig:ppttZCS}
\end{figure}

As it can be seen in Fig.~\ref{fig:ppttZCS} \emph{(left panel)},
and subject to experimental confirmation,
a KK gluon of mass roughly below 1.4 TeV is excluded at $1\sigma$
for $r=0.3$, while for $r=0.8$ the exclusion rules out  
$M\lesssim 1.5\ \textrm{TeV}$.
If one allows for $2\sigma$ deviations, then
only KK-gluon masses below 1 TeV are excluded. 
In addition,
KK gluons of masses 2 and 5 TeV 
are allowed at present \emph{(right panel)} regardless of its widths.
The experimental results from the $pp\to t\bar t Z$ channel are in very good agreement with the SM predictions and the excluded region of 
light KK-gluon masses was already excluded from the more robust results on top-quark pair production in Sec.~\ref{sec:top_pair}. 
It is worth highlighting that the $pp\to t\bar t Z$ results are consistent with a heavy mass KK gluon, 
in contradiction with the analysis of the $pp\to t\bar t W$ cross-section values in Sec.~\ref{ppttZCS}.

\subsection{Top-quark pair plus Higgs production}

\begin{figure}[htb]
    \centering
    \includegraphics[width=12.5cm]{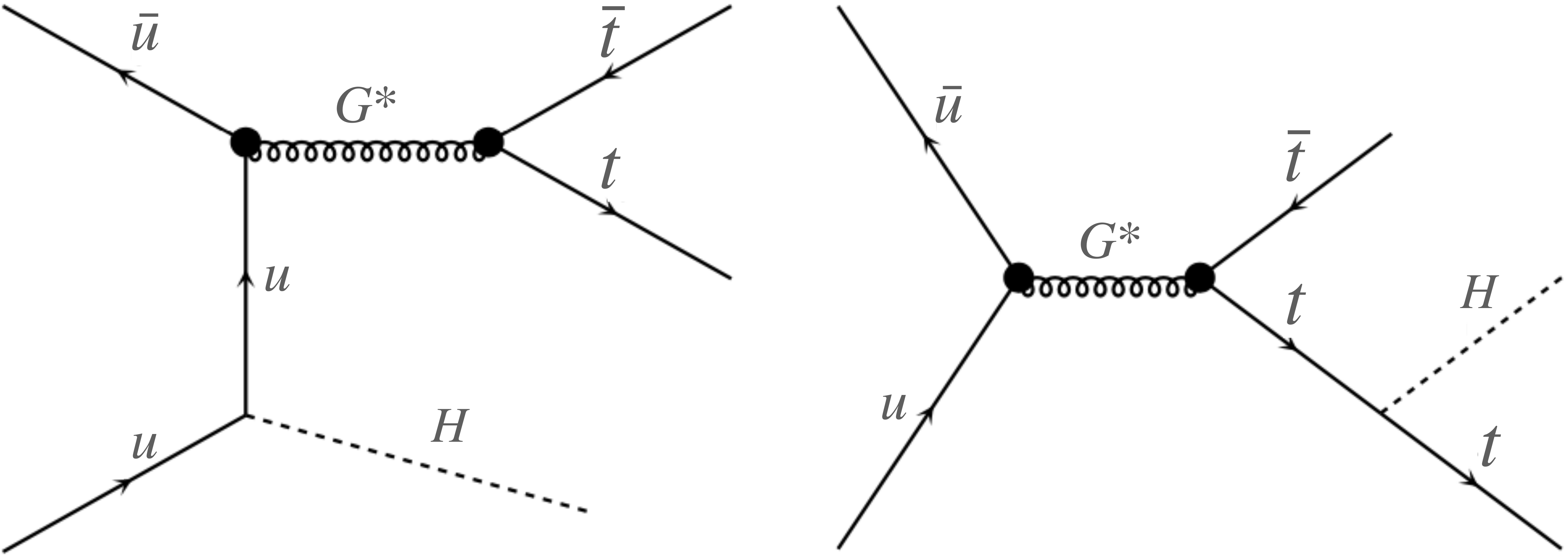}
    \caption{\it A representative subset of Feynman diagrams of the process
    $pp\to t\bar t H$ mediated by a KK gluon.
    These are only two of the 16 diagrams containing a KK gluon
    out of a total of 116 involved in the process.}
    \label{fig:ppttHCSFeynmanD}
\end{figure}

The $pp\to t\bar t H$ total cross-section involves the same 116
Feynman diagrams as the $t\bar t Z$ channel, of course,
replacing the $Z$ by the $H$ (see Fig.~\ref{fig:ppttHCSFeynmanD}).
Similarly, the KK-gluon mass can be reconstructed either from the
$t\bar t$ or the $t\bar t H$ invariant masses.
%The SM cross-section is calculated to be
%$\sigma_{\rm SM}^{t\bar t Z}=0.582\ \textrm{pb}$ in MadGraph5.
The experimental observed ratios are 
$\mu_{t\bar t H}^{\rm obs}=0.70^{+0.36}_{-0.33}$
measured by the ATLAS collaboration \cite{ATLAS:2019nvo} and
$\mu_{t\bar t H}^{\rm obs}=0.92^{+0.25}_{-0.23}$
by the CMS collaboration~\cite{Sirunyan:2020icl}
(both with a reference SM value of
$\sigma_{t\bar t H}^{\rm SM}=507\ \textrm{fb}$).

\begin{figure}[htb]
    \centering
    \includegraphics[width=7.5cm]{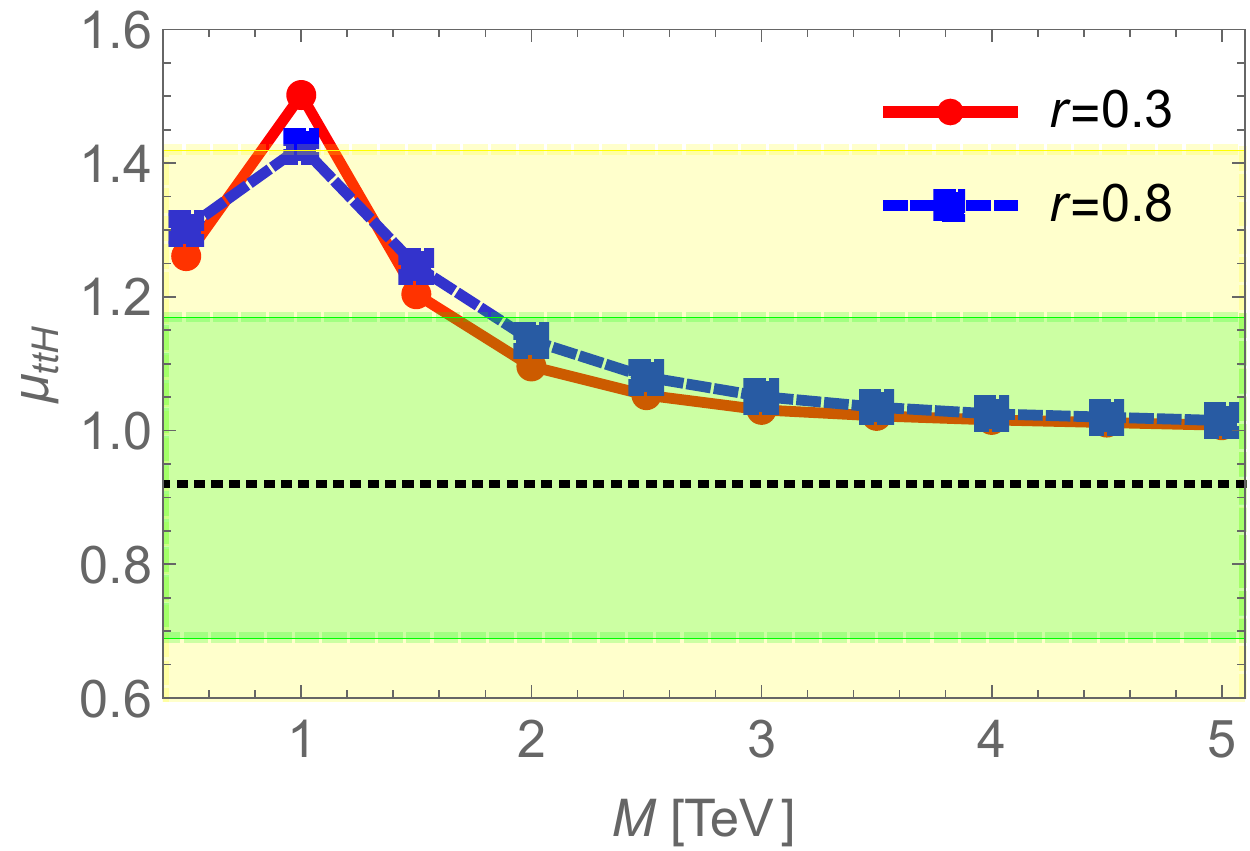}\quad
    \includegraphics[width=7.5cm]{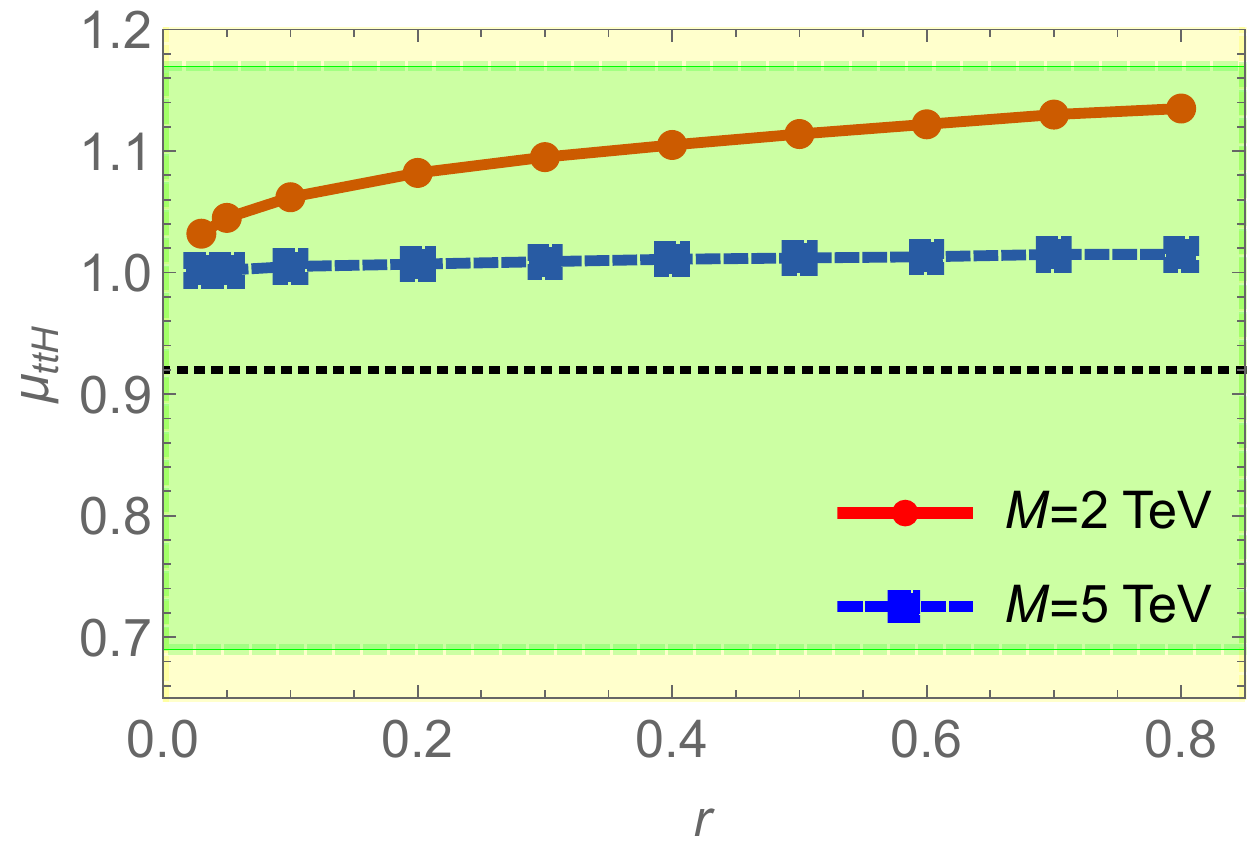}
    \caption{\it Ratio of the total cross-section $pp\to t\bar t H$
    with respect to the SM value, $\mu_{t\bar t H}^{\rm th}$,
    as a function of the renormalized on-shell mass $M$
    \emph{(left panel)} for the fixed ratios
    $r=0.3$ (red solid line) and $r=0.8$ (blue dashed line),
    and as a function of the ratio $r$ \emph{(right panel)}
    for the fixed masses $M=2\ \textrm{TeV}$ (red solid line) and
    $M=5\ \textrm{TeV}$ (blue dashed line).
    The different lines correspond to the predictions based on the
    full propagator (pole approach), while the bands represent
    $1\sigma$ (light green) and $2\sigma$ (light yellow)
    deviations from the $\mu_{t\bar t H}^{\rm obs}$ central value
    (black dotted line) observed by the CMS collaboration \cite{Sirunyan:2020icl}.}
    \label{fig:ppttHCS}
\end{figure}

As is shown in Fig.~\ref{fig:ppttHCS} \emph{(left panel)},
and again subject to experimental verification,
a KK gluon of mass roughly below 1.6 TeV is excluded at $1\sigma$
for $r=0.3$, while for $r=0.8$ any 
$M\lesssim 1.8\ \textrm{TeV}$ is ruled out.
At $2\sigma$, all mass values are allowed with exception of 
$M\simeq 1\ \textrm{TeV}$.
The results from this subsection are consistent with the ones from the analysis of $t\bar t Z$ production 
but in tension with the outcome from $t\bar t W$ production.
%In agreement with the analysis of $t\bar t Z$ production
%but in conflict with the $t\bar t W$ one, 
%KK gluons of masses 2 and 5 TeV disregarding their widths
%are allowed at present \emph{(right panel)}. 
In summary, the production of $pp\to t\bar t H$ is in perfect agreement with the SM predictions, still with large uncertainties, 
and therefore it does not add any relevant information to the possible existence and properties of the heavy KK gluon.

\section{Four top-quark production at the LHC}
\label{ppttttCS}

The four top-quark production at the LHC is, as of now, just a recently
explored area by experimental collaborations~\cite{Aad:2020klt,Sirunyan:2019wxt}
due to its technical difficulties.
This channel offers, in general, a new window for new physics searches
and, in particular, allows for the hunting of a possible existing
KK gluon.
In this section, we compute
the total cross-section $\sigma_{tt\bar t \bar t}=\sigma(pp\to tt\bar t\bar t)$, 
including the SM and KK-gluon contributions,
for different values of the KK-gluon on-shell mass $M$ and
the ratio $r$.
In order to perform an accurate comparison with experimental data,
we use for the SM cross-section the value given in 
Ref.~\cite{Frederix:2017wme}, that is
$\sigma_{tt\bar t\bar t}^{\rm SM}=12\ \textrm{fb}$,
which is calculated at complete next-to-leading order,
and consider it as our reference value.
The contributions of the KK gluon and 
the interference of the KK gluon with the SM,
both computed at leading order, are obtained 
using {\sc MadGraph5$_-$aMC}~\cite{Alwall:2014hca} by
subtracting the SM cross-section from the total cross-section.
Finally, our final predicted values for the total cross-section
are the sum of the reference SM cross-section 
from experiment plus the KK-gluon
and interference cross-sections computed in  {\sc MadGraph5$_-$aMC}
with simulations containing $10^5$ events.

\begin{figure}[htb]
    \centering
    \includegraphics[width=15cm]{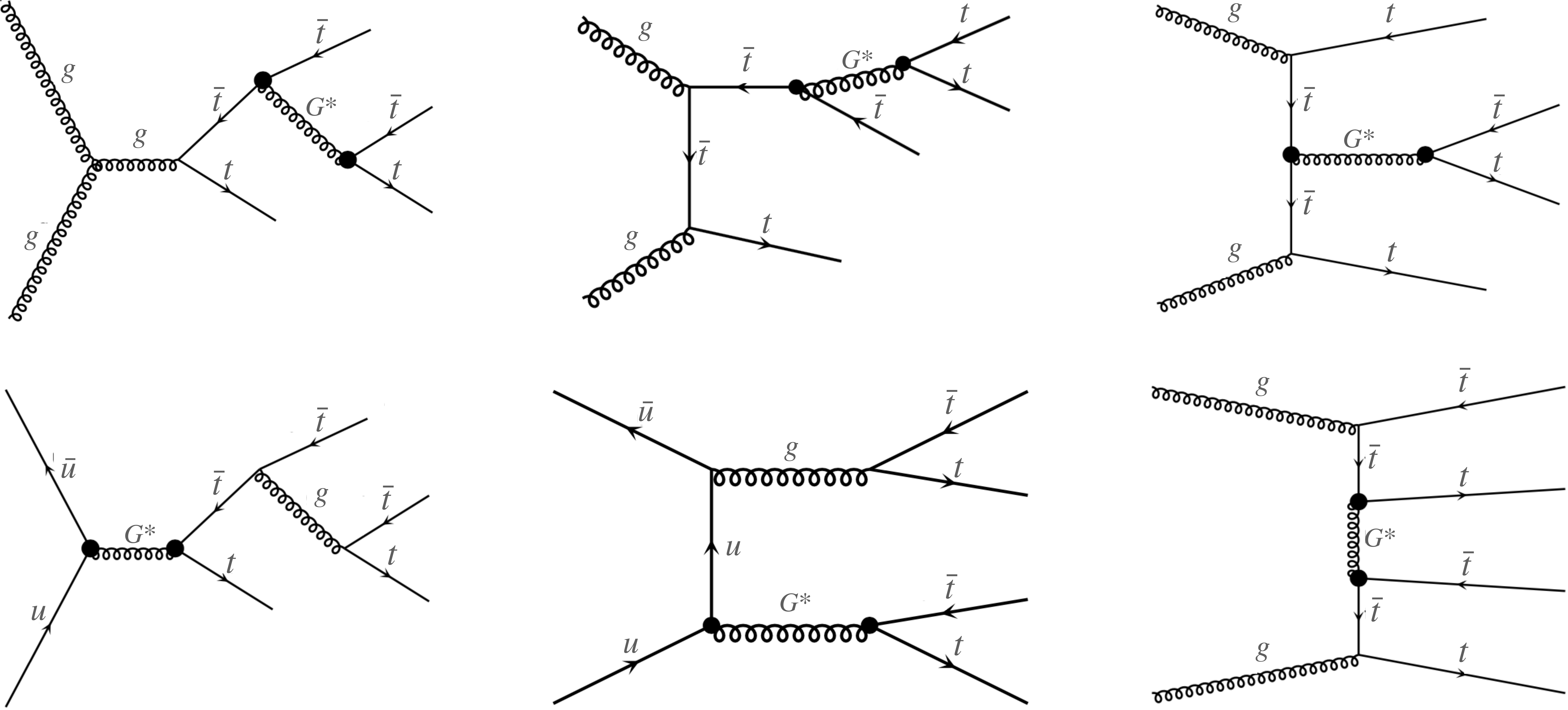}
    \caption{\it A representative subset of Feynman diagrams
    of the process $pp\to tt\bar t\bar t$ mediated by a KK gluon.
    These are only 6 of the 544 diagrams containing a single KK gluon
    out of a total of 1920 involved in the process.}
    \label{fig:ppttttCSFeynmanD}
\end{figure}

The $pp\to tt\bar t\bar t$ total cross-section at leading order
involves a total of 1920 Feynman diagrams,
544 of which involve a KK gluon.
A subset of representative diagrams entering this process
is shown in Fig.~\ref{fig:ppttttCSFeynmanD}.
As was done in the previous section,
we only compare with the experimental data our predictions
based on the use of the full propagator (pole approach).
For this process, we include plots of $\sigma_{tt\bar t\bar t}$
as a function of $M$ for fixed $r=0.3$ and $0.8$, and 
as a function of $r$ for fixed 
$M=2\ \textrm{TeV}$ and 5 TeV.
Our predictions are shown in red (solid) and blue (dashed) lines,
%(depending on the value of $M$ or $r$),
while the observed cross-section is displayed using bands 
corresponding to $1\sigma$ (light green) and $2\sigma$ (light yellow)
deviations from the $\sigma_{tt\bar t\bar t}^{\rm obs}$
(black dotted line) central value.
From the two experimental cross-sections, 
$\sigma_{tt\bar t\bar t}^{\rm obs}=24^{+7}_{-6}\ \textrm{fb}$
measured by the ATLAS collaboration \cite{Aad:2020klt} and
$\sigma_{tt\bar t\bar t}^{\rm obs}=9.4^{+6.2}_{-5.6}\ \textrm{fb}$
by the CMS collaboration~\cite{Sirunyan:2019wxt},
we choose for comparison the latter as it is closer to 
the reference SM cross-section.

\begin{figure}[htb]
    \centering
    \includegraphics[width=7.5cm]{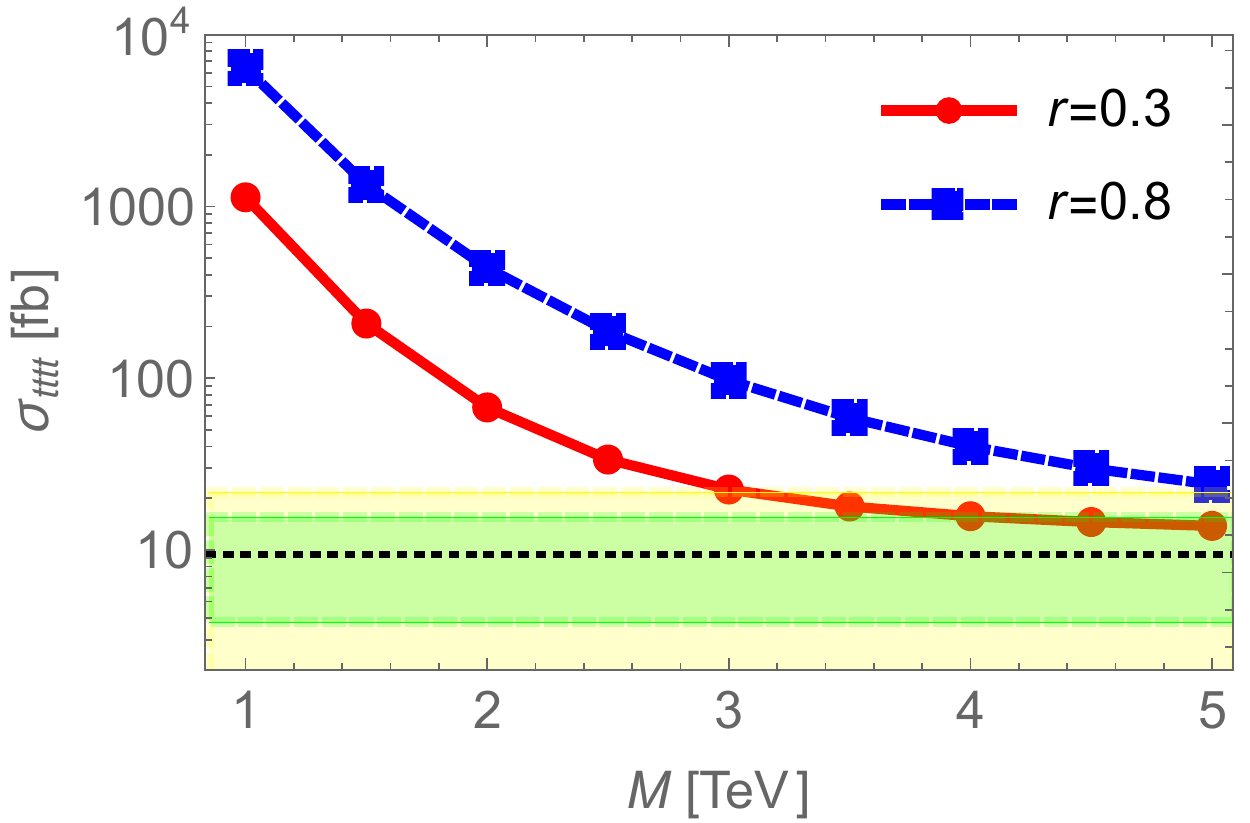}\quad
    \includegraphics[width=7.5cm]{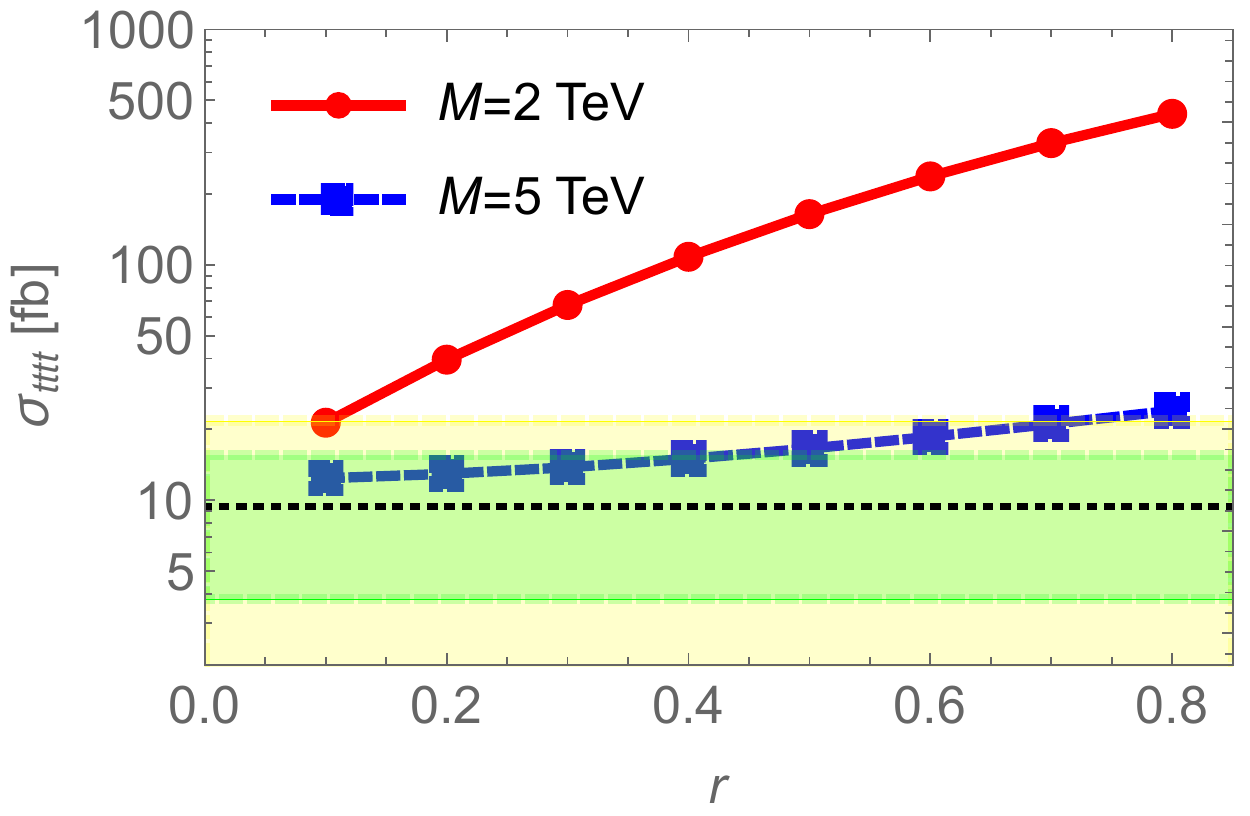}
    \caption{\it The total cross-section $pp\to tt\bar t\bar t$
    as a function of the renormalized on-shell mass $M$
    \emph{(left panel)} for the fixed ratios
    $r=0.3$ (red solid line) and $r=0.8$ (blue dashed line),
    and as a function of the ratio $r$ \emph{(right panel)}
    for the fixed masses $M=2\ \textrm{TeV}$ (red solid line) and
    $M=5\ \textrm{TeV}$ (blue dashed line).
    The different lines correspond to the predictions based on the
    full propagator (pole approach), while the bands represent
    $1\sigma$ (light green) and $2\sigma$ (light yellow)
    deviations from the $\sigma_{tt\bar t\bar t}^{\rm obs}$
    central value (black dotted line) observed by the 
    CMS collaboration \cite{Sirunyan:2019wxt}.}
    \label{fig:ppttttCS}
\end{figure}

In Fig.~\ref{fig:ppttttCS} \emph{(left panel)}, 
the total $pp\to tt\bar t\bar t$ cross-section is shown
as a function of $M$ for $r=0.3$ (red solid line) and 
$r=0.8$ (blue dashed line).
As it can be seen, a KK gluon of $M\lesssim 4\ \textrm{TeV}$
is excluded for $r=0.3$ at the $1\sigma$ CL, whereas for 
$r=0.8$ all the values of $M$ are excluded.
At $2\sigma$, the exclusion limits are relaxed to
$M\lesssim 2.8\ \textrm{TeV}$ for $r=0.3$ and
$M\lesssim 4.6\ \textrm{TeV}$ for $r=0.8$.
In the first case, the exclusion limit seems to be in reasonable
agreement with the one obtained from the top-quark pair cross-section,
$M\lesssim 3.5\ \textrm{TeV}$ for $r=0.3$. 
The right panel of Fig.~\ref{fig:ppttttCS} %\emph{(right panel)}, 
displays the total $pp\to tt\bar t\bar t$ cross-section as a function of $r$
for $M=2\ \textrm{TeV}$ (red solid line) and 
$M=5\ \textrm{TeV}$ (blue dashed line).
Clearly, a KK gluon of $M=2\ \textrm{TeV}$
is excluded regardless of its width at the $1\sigma$ CL,
while for $M=5\ \textrm{TeV}$ the exclusion limit rules out any
$r\gtrsim 0.4$ or $\Gamma\gtrsim 2\ \textrm{TeV}$.
At the $2\sigma$ level, only a very narrow KK gluon
with $r\simeq 0.1$ or $\Gamma\simeq 200\ \textrm{GeV}$ 
is allowed for the $M=2\ \textrm{TeV}$ case,
while for the $M=5\ \textrm{TeV}$ case the exclusion limits discard any 
$r\gtrsim 0.7$ or $\Gamma\gtrsim 3.5\ \textrm{TeV}$.

In summary, the $pp\to t t\bar t \bar t$ channel seems the most promising one, as the SM prediction and signal are of similar size. 
For the moment, though, our simplified analysis appears to exclude, at the $2\sigma$ level, resonances with $M\lesssim4.6$ TeV for $r=0.8$, 
and with $r\gtrsim0.7$ for $M=5$ TeV, both regions being allowed by the present top-pair production experimental data (cf.~Sec.~\ref{sec:top_pair}). 
It appears from the present analysis that future data on the $pp\to t t\bar t \bar t$ channel shall be most relevant for uncovering new physics and, 
in particular, for detecting the presence of heavy KK gluons, a remnant of warped theories aiming to solve the hierarchy problem.

\section{Conclusions}
\label{sec:conclusions}

In theories with a warped extra dimension and two branes, i.e.~the UV and the IR, 
the Higgs field must be localized toward the IR brane in order to solve the hierarchy problem. 
In these theories, where all Standard Model fields are propagating in the bulk of the extra dimension, 
Kaluza-Klein excitations are also localized toward the IR brane and interact strongly with all IR localized fields, 
or composite fields in the language of the dual 4D holographic theory. 
Moreover, the flavor problem in the quark sector can be solved if heavy quarks are very composite, and light quarks elementary. 
In this way, the most composite fermion ought to be the right-handed top quark $t_{R}$, 
which then can interact very strongly with the Kaluza-Klein states. 
Out of these Kaluza-Klein states, the gluons are the most model independent ones, 
as they do not interfere with the electroweak symmetry breaking phenomenon, and thus its phenomenology is, to a very large extent, model independent. 
Accordingly, the Kaluza-Klein gluon can decay mainly into the top-quark pair and get a broad width, 
which can have consequences at the phenomenological and experimental level.

In this work, we have studied the case of the first Kaluza-Klein gluon resonance with a width that mainly depends on the coupling to the quark $t_R$, 
which, in turn, depends on its particular localization in the extra dimension. 
We have found that the strong coupling with the quark sector has two effects: on one hand, as we have described, 
it can produce a broad resonance which can easily jeopardize experimental capabilities of detection; 
on the other hand, it can have a strong renormalization effect on its mass, which can also strongly change its production rate at LHC. 
Both effects are taken into account by computing the pole mass $M_{\rm p}$ and pole width $\Gamma_{\rm p}$ 
as the full propagator pole in the second Riemann sheet of the complex $s$-plane, defined as $s_{\rm p}=M_{\rm p}^2-i M_{\rm p} \Gamma_{\rm p}$. 
We have then compared, for the $t\bar t$ channel, the predictions of the full propagator with that of the commonly used Breit-Wigner approximation, 
using renormalized values for the mass $M$ and width $\Gamma$. Admittedly, the latter is relatively accurate only for the case of narrow resonances.   

The comparison of pole and renormalized quantities yield meaningful differences, depending, of course, 
on the strength of the coupling of the Kaluza-Klein gluon to the quark $t_R$. The main qualitative results are:
\begin{itemize}
\item
The pole mass $M_{\rm p}$ is smaller than the renormalized mass $M$ by as much as 40\%. 
This has a great impact in the production cross-sections for the different processes we have analyzed.
\item
For small couplings, the pole width $\Gamma_{\rm p}$ is of the order of the renormalized one $\Gamma$; 
however, for large couplings, $\Gamma_{\rm p}$ is smaller than $\Gamma$ by as much as 30\%.
\end{itemize}

We have computed with {\sc MadGraph5$_-$aMC} 
the cross-section of the $pp\to t\bar t$ process using the full propagator and the Breit-Wigner approximation, and compared the results. 
This has been done for different values of the renormalized mass $1 \textrm{ TeV}<M<5 \textrm{ TeV}$ and width $\Gamma/M<0.8$. 
As well as this, a comparison between our theoretical results for this channel with the available experimental data from the ATLAS collaboration 
has been performed. 
In addition, we have calculated cross-sections for the processes $t\bar t W$, $t\bar t Z$, $t\bar t H$, and $t t \bar t \bar t$, and, again, 
compared with available empirical data from the ATLAS and CMS collaborations.
In particular, we have found differences of order 100\% for the case of broad resonances in the $t\bar t$ channel cross-sections, 
depending on whether the full or the Breit-Wigner propagator was employed. 
As for the $t\bar t W$, $t\bar t Z$, $t\bar t H$, and $t t \bar t \bar t$ channels, 
given that the experimental data are not accurate enough at present, we have just presented results using the full propagator for completeness. 
Once again, the theoretical predictions depend dramatically on the width of the resonance. 
We have found particularly interesting the $t t \bar t \bar t$ channel for which the Standard Model prediction and the corrections from 
Kaluza-Klein gluons are essentially of the same order of magnitude. 
Once again, the theoretical predictions in this channel can vary by more than one order of magnitude between narrow and large resonances. 

Broader resonances are more strongly coupled (in this model, the strength of the coupling is related to the width of the resonance) 
and pole masses associated to these resonances turn out to be significantly smaller than the corresponding renormalized ones, 
which, in turn, implies that they are more copiously produced. This has a remarkable impact on the experimental signature. 
Unfortunately, at present, experimental searches do not cover the case of very broad resonances. 
We hope, however, that future experimental endeavors will consider this scenario so that strong experimental bounds on resonance masses and widths 
can be extracted, and robust comparison with the presently analyzed theoretical model's predictions can be performed. 
This will decisively point out the capabilities and limitations of the theory.

\vspace{0.5cm}
\section*{Acknowledgments}
We would like to thank A.~Juste and M.~Szewc for having participated in the early stages of this work, and E.~Megias for discussions. 
This work is supported by the Secretaria d'Universitats i Recerca del Departament d'Empresa i Coneixement de la Generalitat de Catalunya 
under the grant 2017SGR1069, by the Ministerio de Econom\'{i}a, Industria y Competitividad under grants FPA2017-88915-P and FPA2017-86989-P, 
from the Centro de Excelencia Severo Ochoa under the grant SEV-2016-0588 and 
from the EU STRONG-2020 project under the program H2020-INFRAIA-2018-1, grant agreement No. 824093. 
IFAE is partially funded by the CERCA program of the Generalitat de Catalunya.

\bibliographystyle{JHEP}
\bibliography{refs}

\providecommand{\href}[2]{#2}\begingroup\raggedright\begin{thebibliography}{10}

\bibitem{Martin:1997ns}
S.~P. Martin, \emph{{A Supersymmetry primer}},
  \href{http://dx.doi.org/10.1142/9789812839657_0001}{\emph{Adv. Ser. Direct.
  High Energy Phys.} {\bfseries 21} (2010) 1--153},
  [\href{https://arxiv.org/abs/hep-ph/9709356}{{\ttfamily hep-ph/9709356}}].

\bibitem{Randall:1999ee}
L.~Randall and R.~Sundrum, \emph{{A Large mass hierarchy from a small extra
  dimension}}, \href{http://dx.doi.org/10.1103/PhysRevLett.83.3370}{\emph{Phys.
  Rev. Lett.} {\bfseries 83} (1999) 3370--3373},
  [\href{https://arxiv.org/abs/hep-ph/9905221}{{\ttfamily hep-ph/9905221}}].

\bibitem{Stancato:2008mp}
D.~Stancato and J.~Terning, \emph{{The Unhiggs}},
  \href{http://dx.doi.org/10.1088/1126-6708/2009/11/101}{\emph{JHEP} {\bfseries
  11} (2009) 101}, [\href{https://arxiv.org/abs/0807.3961}{{\ttfamily
  0807.3961}}].

\bibitem{Stancato:2010ay}
D.~Stancato and J.~Terning, \emph{{Constraints on the Unhiggs Model from Top
  Quark Decay}},
  \href{http://dx.doi.org/10.1103/PhysRevD.81.115012}{\emph{Phys. Rev.}
  {\bfseries D81} (2010) 115012},
  [\href{https://arxiv.org/abs/1002.1694}{{\ttfamily 1002.1694}}].

\bibitem{Falkowski:2008yr}
A.~Falkowski and M.~Perez-Victoria, \emph{{Holographic Unhiggs}},
  \href{http://dx.doi.org/10.1103/PhysRevD.79.035005}{\emph{Phys. Rev.}
  {\bfseries D79} (2009) 035005},
  [\href{https://arxiv.org/abs/0810.4940}{{\ttfamily 0810.4940}}].

\bibitem{Cabrer:2009we}
J.~A. Cabrer, G.~von Gersdorff and M.~Quiros, \emph{{Soft-Wall Stabilization}},
  \href{http://dx.doi.org/10.1088/1367-2630/12/7/075012}{\emph{New J. Phys.}
  {\bfseries 12} (2010) 075012},
  [\href{https://arxiv.org/abs/0907.5361}{{\ttfamily 0907.5361}}].

\bibitem{Englert:2012dq}
C.~Englert, M.~Spannowsky, D.~Stancato and J.~Terning, \emph{{Unconstraining
  the Unhiggs}},
  \href{http://dx.doi.org/10.1103/PhysRevD.85.095003}{\emph{Phys. Rev.}
  {\bfseries D85} (2012) 095003},
  [\href{https://arxiv.org/abs/1203.0312}{{\ttfamily 1203.0312}}].

\bibitem{Englert:2012cb}
C.~Englert, D.~G. Netto, M.~Spannowsky and J.~Terning, \emph{{Constraining the
  Unhiggs with LHC data}},
  \href{http://dx.doi.org/10.1103/PhysRevD.86.035010}{\emph{Phys. Rev.}
  {\bfseries D86} (2012) 035010},
  [\href{https://arxiv.org/abs/1205.0836}{{\ttfamily 1205.0836}}].

\bibitem{Goncalves:2018pkt}
D.~Gonzalves, T.~Han and S.~Mukhopadhyay, \emph{{Higgs Couplings at High
  Scales}}, \href{http://dx.doi.org/10.1103/PhysRevD.98.015023}{\emph{Phys.
  Rev.} {\bfseries D98} (2018) 015023},
  [\href{https://arxiv.org/abs/1803.09751}{{\ttfamily 1803.09751}}].

\bibitem{Csaki:2018kxb}
C.~Csaki, G.~Lee, S.~J. Lee, S.~Lombardo and O.~Telem, \emph{{Continuum
  Naturalness}}, \href{http://dx.doi.org/10.1007/JHEP03(2019)142}{\emph{JHEP}
  {\bfseries 03} (2019) 142},
  [\href{https://arxiv.org/abs/1811.06019}{{\ttfamily 1811.06019}}].

\bibitem{Lee:2018fxj}
S.~J. Lee, M.~Park and Z.~Qian, \emph{{Probing unitarity violation in the tail
  of the off-shell Higgs boson in $V_LV_L$ mode}},
  \href{http://dx.doi.org/10.1103/PhysRevD.100.011702}{\emph{Phys. Rev.}
  {\bfseries D100} (2019) 011702},
  [\href{https://arxiv.org/abs/1812.02679}{{\ttfamily 1812.02679}}].

\bibitem{Megias:2019vdb}
E.~Megias and M.~Quiros, \emph{{Gapped Continuum Kaluza-Klein spectrum}},
  \href{http://dx.doi.org/10.1007/JHEP08(2019)166}{\emph{JHEP} {\bfseries 08}
  (2019) 166}, [\href{https://arxiv.org/abs/1905.07364}{{\ttfamily
  1905.07364}}].

\bibitem{Shirazi:2019bjw}
A.~Shayegan~Shirazi and J.~Terning, \emph{{Quantum Critical Higgs: from
  AdS$_{5}$ to colliders}},
  \href{http://dx.doi.org/10.1007/JHEP02(2020)026}{\emph{JHEP} {\bfseries 02}
  (2020) 026}, [\href{https://arxiv.org/abs/1908.06186}{{\ttfamily
  1908.06186}}].

\bibitem{Gao:2019gfw}
C.~Gao, A.~Shayegan~Shirazi and J.~Terning, \emph{{Collider Phenomenology of a
  Gluino Continuum}},
  \href{http://dx.doi.org/10.1007/JHEP01(2020)102}{\emph{JHEP} {\bfseries 01}
  (2020) 102}, [\href{https://arxiv.org/abs/1909.04061}{{\ttfamily
  1909.04061}}].

\bibitem{Georgi:2007ek}
H.~Georgi, \emph{{Unparticle physics}},
  \href{http://dx.doi.org/10.1103/PhysRevLett.98.221601}{\emph{Phys. Rev.
  Lett.} {\bfseries 98} (2007) 221601},
  [\href{https://arxiv.org/abs/hep-ph/0703260}{{\ttfamily hep-ph/0703260}}].

\bibitem{Alwall:2014hca}
J.~Alwall, R.~Frederix, S.~Frixione, V.~Hirschi, F.~Maltoni, O.~Mattelaer
  et~al., \emph{{The automated computation of tree-level and next-to-leading
  order differential cross sections, and their matching to parton shower
  simulations}}, \href{http://dx.doi.org/10.1007/JHEP07(2014)079}{\emph{JHEP}
  {\bfseries 07} (2014) 079},
  [\href{https://arxiv.org/abs/1405.0301}{{\ttfamily 1405.0301}}].

\bibitem{Agashe:2006hk}
K.~Agashe, A.~Belyaev, T.~Krupovnickas, G.~Perez and J.~Virzi, \emph{{LHC
  Signals from Warped Extra Dimensions}},
  \href{http://dx.doi.org/10.1103/PhysRevD.77.015003}{\emph{Phys. Rev. D}
  {\bfseries 77} (2008) 015003},
  [\href{https://arxiv.org/abs/hep-ph/0612015}{{\ttfamily hep-ph/0612015}}].

\bibitem{Lillie:2007yh}
B.~Lillie, L.~Randall and L.-T. Wang, \emph{{The Bulk RS KK-gluon at the LHC}},
  \href{http://dx.doi.org/10.1088/1126-6708/2007/09/074}{\emph{JHEP} {\bfseries
  09} (2007) 074}, [\href{https://arxiv.org/abs/hep-ph/0701166}{{\ttfamily
  hep-ph/0701166}}].

\bibitem{Barcelo:2011fw}
R.~Barcelo, A.~Carmona, M.~Masip and J.~Santiago, \emph{{Gluon excitations in t
  tbar production at hadron colliders}},
  \href{http://dx.doi.org/10.1103/PhysRevD.84.014024}{\emph{Phys. Rev.}
  {\bfseries D84} (2011) 014024},
  [\href{https://arxiv.org/abs/1105.3333}{{\ttfamily 1105.3333}}].

\bibitem{Barcelo:2011vk}
R.~Barcelo, A.~Carmona, M.~Masip and J.~Santiago, \emph{{Stealth gluons at
  hadron colliders}},
  \href{http://dx.doi.org/10.1016/j.physletb.2011.12.002}{\emph{Phys. Lett.}
  {\bfseries B707} (2012) 88--91},
  [\href{https://arxiv.org/abs/1106.4054}{{\ttfamily 1106.4054}}].

\bibitem{Dasgupta:2019yjm}
S.~Dasgupta, S.~K. Rai and T.~S. Ray, \emph{{Impact of a colored vector
  resonance on the collider constraints for top-like top partner}},
  \href{http://dx.doi.org/10.1103/PhysRevD.102.115014}{\emph{Phys. Rev. D}
  {\bfseries 102} (2020) 115014},
  [\href{https://arxiv.org/abs/1912.13022}{{\ttfamily 1912.13022}}].

\bibitem{Liu:2019bua}
D.~Liu, L.-T. Wang and K.-P. Xie, \emph{{Broad composite resonances and their
  signals at the LHC}},
  \href{http://dx.doi.org/10.1103/PhysRevD.100.075021}{\emph{Phys. Rev. D}
  {\bfseries 100} (2019) 075021},
  [\href{https://arxiv.org/abs/1901.01674}{{\ttfamily 1901.01674}}].

\bibitem{Jung:2019iii}
S.~Jung, D.~Lee and K.-P. Xie, \emph{{Beyond $M_{t\bar{t}}$: learning to search
  for a broad $t\bar t$ resonance at the LHC}},
  \href{http://dx.doi.org/10.1140/epjc/s10052-020-7672-9}{\emph{Eur. Phys. J.
  C} {\bfseries 80} (2020) 105},
  [\href{https://arxiv.org/abs/1906.02810}{{\ttfamily 1906.02810}}].

\bibitem{Sirunyan:2019wxt}
{\scshape CMS} collaboration, A.~M. Sirunyan et~al., \emph{{Search for
  production of four top quarks in final states with same-sign or multiple
  leptons in proton-proton collisions at $\sqrt{s}=$ 13 TeV}},
  \href{http://dx.doi.org/10.1140/epjc/s10052-019-7593-7}{\emph{Eur. Phys. J.
  C} {\bfseries 80} (2020) 75},
  [\href{https://arxiv.org/abs/1908.06463}{{\ttfamily 1908.06463}}].

\bibitem{Aad:2020klt}
{\scshape ATLAS} collaboration, G.~Aad et~al., \emph{{Evidence for
  $t\bar{t}t\bar{t}$ production in the multilepton final state in
  proton\textendash{}proton collisions at $\sqrt{s}=13$ $\text {TeV}$ with the
  ATLAS detector}},
  \href{http://dx.doi.org/10.1140/epjc/s10052-020-08509-3}{\emph{Eur. Phys. J.
  C} {\bfseries 80} (2020) 1085},
  [\href{https://arxiv.org/abs/2007.14858}{{\ttfamily 2007.14858}}].

\bibitem{Sirunyan:2020icl}
{\scshape CMS} collaboration, A.~M. Sirunyan et~al., \emph{{Measurement of the
  Higgs boson production rate in association with top quarks in final states
  with electrons, muons, and hadronically decaying tau leptons at $\sqrt{s} =$
  13 TeV}},  \href{https://arxiv.org/abs/2011.03652}{{\ttfamily 2011.03652}}.

\bibitem{ATLAS:2019nvo}
{\scshape ATLAS} collaboration, \emph{{Analysis of $t\bar{t}H$ and $t\bar{t}W$
  production in multilepton final states with the ATLAS detector}},
  \href{https://arxiv.org/abs/ATLAS-CONF-2019-045}{{\ttfamily
  ATLAS-CONF-2019-045}}.

\bibitem{CMS:2019too}
{\scshape CMS} collaboration, A.~M. Sirunyan et~al., \emph{{Measurement of top
  quark pair production in association with a Z boson in proton-proton
  collisions at $\sqrt{s}=$ 13 TeV}},
  \href{http://dx.doi.org/10.1007/JHEP03(2020)056}{\emph{JHEP} {\bfseries 03}
  (2020) 056}, [\href{https://arxiv.org/abs/1907.11270}{{\ttfamily
  1907.11270}}].

\bibitem{ATLAS:2020cxf}
{\scshape ATLAS} collaboration, \emph{{Measurements of the inclusive and
  differential production cross sections of a top-quark-antiquark pair in
  association with a $Z$ boson at $\sqrt{s} = 13$ TeV with the ATLAS
  detector}},  \href{https://arxiv.org/abs/ATLAS-CONF-2020-028}{{\ttfamily
  ATLAS-CONF-2020-028}}.

\bibitem{Goldberger:1999uk}
W.~D. Goldberger and M.~B. Wise, \emph{{Modulus stabilization with bulk
  fields}}, \href{http://dx.doi.org/10.1103/PhysRevLett.83.4922}{\emph{Phys.
  Rev. Lett.} {\bfseries 83} (1999) 4922--4925},
  [\href{https://arxiv.org/abs/hep-ph/9907447}{{\ttfamily hep-ph/9907447}}].

\bibitem{DeWolfe:1999cp}
O.~DeWolfe, D.~Z. Freedman, S.~S. Gubser and A.~Karch, \emph{{Modeling the
  fifth-dimension with scalars and gravity}},
  \href{http://dx.doi.org/10.1103/PhysRevD.62.046008}{\emph{Phys. Rev.}
  {\bfseries D62} (2000) 046008},
  [\href{https://arxiv.org/abs/hep-th/9909134}{{\ttfamily hep-th/9909134}}].

\bibitem{Agashe:2003zs}
K.~Agashe, A.~Delgado, M.~J. May and R.~Sundrum, \emph{{RS1, custodial isospin
  and precision tests}},
  \href{http://dx.doi.org/10.1088/1126-6708/2003/08/050}{\emph{JHEP} {\bfseries
  08} (2003) 050}, [\href{https://arxiv.org/abs/hep-ph/0308036}{{\ttfamily
  hep-ph/0308036}}].

\bibitem{Carena:2018cow}
M.~Carena, E.~Megias, M.~Quiros and C.~Wagner, \emph{{$
  {R}_{D^{\left(*\right)}} $ in custodial warped space}},
  \href{http://dx.doi.org/10.1007/JHEP12(2018)043}{\emph{JHEP} {\bfseries 12}
  (2018) 043}, [\href{https://arxiv.org/abs/1809.01107}{{\ttfamily
  1809.01107}}].

\bibitem{Contino:2003ve}
R.~Contino, Y.~Nomura and A.~Pomarol, \emph{{Higgs as a holographic
  pseudoGoldstone boson}},
  \href{http://dx.doi.org/10.1016/j.nuclphysb.2003.08.027}{\emph{Nucl. Phys.}
  {\bfseries B671} (2003) 148--174},
  [\href{https://arxiv.org/abs/hep-ph/0306259}{{\ttfamily hep-ph/0306259}}].

\bibitem{Agashe:2004rs}
K.~Agashe, R.~Contino and A.~Pomarol, \emph{{The Minimal composite Higgs
  model}}, \href{http://dx.doi.org/10.1016/j.nuclphysb.2005.04.035}{\emph{Nucl.
  Phys.} {\bfseries B719} (2005) 165--187},
  [\href{https://arxiv.org/abs/hep-ph/0412089}{{\ttfamily hep-ph/0412089}}].

\bibitem{Contino:2006qr}
R.~Contino, L.~Da~Rold and A.~Pomarol, \emph{{Light custodians in natural
  composite Higgs models}},
  \href{http://dx.doi.org/10.1103/PhysRevD.75.055014}{\emph{Phys. Rev.}
  {\bfseries D75} (2007) 055014},
  [\href{https://arxiv.org/abs/hep-ph/0612048}{{\ttfamily hep-ph/0612048}}].

\bibitem{Cabrer:2010si}
J.~A. Cabrer, G.~von Gersdorff and M.~Quiros, \emph{{Warped Electroweak
  Breaking Without Custodial Symmetry}},
  \href{http://dx.doi.org/10.1016/j.physletb.2011.01.058}{\emph{Phys. Lett.}
  {\bfseries B697} (2011) 208--214},
  [\href{https://arxiv.org/abs/1011.2205}{{\ttfamily 1011.2205}}].

\bibitem{Cabrer:2011fb}
J.~A. Cabrer, G.~von Gersdorff and M.~Quiros, \emph{{Suppressing Electroweak
  Precision Observables in 5D Warped Models}},
  \href{http://dx.doi.org/10.1007/JHEP05(2011)083}{\emph{JHEP} {\bfseries 05}
  (2011) 083}, [\href{https://arxiv.org/abs/1103.1388}{{\ttfamily 1103.1388}}].

\bibitem{Cabrer:2011vu}
J.~A. Cabrer, G.~von Gersdorff and M.~Quiros, \emph{{Improving Naturalness in
  Warped Models with a Heavy Bulk Higgs Boson}},
  \href{http://dx.doi.org/10.1103/PhysRevD.84.035024}{\emph{Phys. Rev.}
  {\bfseries D84} (2011) 035024},
  [\href{https://arxiv.org/abs/1104.3149}{{\ttfamily 1104.3149}}].

\bibitem{Carmona:2011ib}
A.~Carmona, E.~Ponton and J.~Santiago, \emph{{Phenomenology of Non-Custodial
  Warped Models}}, \href{http://dx.doi.org/10.1007/JHEP10(2011)137}{\emph{JHEP}
  {\bfseries 10} (2011) 137},
  [\href{https://arxiv.org/abs/1107.1500}{{\ttfamily 1107.1500}}].

\bibitem{Cabrer:2011qb}
J.~A. Cabrer, G.~von Gersdorff and M.~Quiros, \emph{{Flavor Phenomenology in
  General 5D Warped Spaces}},
  \href{http://dx.doi.org/10.1007/JHEP01(2012)033}{\emph{JHEP} {\bfseries 01}
  (2012) 033}, [\href{https://arxiv.org/abs/1110.3324}{{\ttfamily 1110.3324}}].

\bibitem{Quiros:2013yaa}
M.~Quiros, \emph{{Higgs Bosons in Extra Dimensions}},
  \href{http://dx.doi.org/10.1142/S021773231540012X}{\emph{Mod. Phys. Lett.}
  {\bfseries A30} (2015) 1540012},
  [\href{https://arxiv.org/abs/1311.2824}{{\ttfamily 1311.2824}}].

\bibitem{deBlas:2012qf}
J.~de~Blas, A.~Delgado, B.~Ostdiek and A.~de~la Puente, \emph{{LHC Signals of
  Non-Custodial Warped 5D Models}},
  \href{http://dx.doi.org/10.1103/PhysRevD.86.015028}{\emph{Phys. Rev.}
  {\bfseries D86} (2012) 015028},
  [\href{https://arxiv.org/abs/1206.0699}{{\ttfamily 1206.0699}}].

\bibitem{Megias:2015ory}
E.~Megias, O.~Pujolas and M.~Quiros, \emph{{On dilatons and the LHC diphoton
  excess}}, \href{http://dx.doi.org/10.1007/JHEP05(2016)137}{\emph{JHEP}
  {\bfseries 05} (2016) 137},
  [\href{https://arxiv.org/abs/1512.06106}{{\ttfamily 1512.06106}}].

\bibitem{Megias:2016bde}
E.~Megias, G.~Panico, O.~Pujolas and M.~Quiros, \emph{{A Natural origin for the
  LHCb anomalies}},
  \href{http://dx.doi.org/10.1007/JHEP09(2016)118}{\emph{JHEP} {\bfseries 09}
  (2016) 118}, [\href{https://arxiv.org/abs/1608.02362}{{\ttfamily
  1608.02362}}].

\bibitem{Megias:2017ove}
E.~Megias, M.~Quiros and L.~Salas, \emph{{Lepton-flavor universality violation
  in R$_{K}$ and $ {R}_{D^{{\left(\ast \right)}}} $ from warped space}},
  \href{http://dx.doi.org/10.1007/JHEP07(2017)102}{\emph{JHEP} {\bfseries 07}
  (2017) 102}, [\href{https://arxiv.org/abs/1703.06019}{{\ttfamily
  1703.06019}}].

\bibitem{Falkowski:2017pss}
A.~Falkowski, M.~Gonzalez-Alonso and K.~Mimouni, \emph{{Compilation of
  low-energy constraints on 4-fermion operators in the SMEFT}},
  \href{http://dx.doi.org/10.1007/JHEP08(2017)123}{\emph{JHEP} {\bfseries 08}
  (2017) 123}, [\href{https://arxiv.org/abs/1706.03783}{{\ttfamily
  1706.03783}}].

\bibitem{Willenbrock:1991hu}
S.~Willenbrock and G.~Valencia, \emph{{On the definition of the Z boson mass}},
  \href{http://dx.doi.org/10.1016/0370-2693(91)90843-F}{\emph{Phys. Lett. B}
  {\bfseries 259} (1991) 373--376}.

\bibitem{Bhattacharya:1991gr}
T.~Bhattacharya and S.~Willenbrock, \emph{{Particles near threshold}},
  \href{http://dx.doi.org/10.1103/PhysRevD.47.4022}{\emph{Phys. Rev. D}
  {\bfseries 47} (1993) 4022--4027}.

\bibitem{Escribano:2002iv}
R.~Escribano, A.~Gallegos, J.~L. Lucio~M, G.~Moreno and J.~Pestieau, \emph{{On
  the mass, width and coupling constants of the f(0)(980)}},
  \href{http://dx.doi.org/10.1140/epjc/s2003-01140-6}{\emph{Eur. Phys. J. C}
  {\bfseries 28} (2003) 107--114},
  [\href{https://arxiv.org/abs/hep-ph/0204338}{{\ttfamily hep-ph/0204338}}].

\bibitem{Eichten:1984eu}
E.~Eichten, I.~Hinchliffe, K.~D. Lane and C.~Quigg, \emph{{Super Collider
  Physics}}, \href{http://dx.doi.org/10.1103/RevModPhys.56.579,
  10.1103/RevModPhys.58.1065}{\emph{Rev. Mod. Phys.} {\bfseries 56} (1984)
  579--707}.

\bibitem{Aaboud:2019roo}
{\scshape ATLAS} collaboration, M.~Aaboud et~al., \emph{{Search for heavy
  particles decaying into a top-quark pair in the fully hadronic final state in
  $pp$ collisions at $\sqrt{s} =$ 13 TeV with the ATLAS detector}},
  \href{http://dx.doi.org/10.1103/PhysRevD.99.092004}{\emph{Phys. Rev.}
  {\bfseries D99} (2019) 092004},
  [\href{https://arxiv.org/abs/1902.10077}{{\ttfamily 1902.10077}}].

\bibitem{Frederix:2017wme}
R.~Frederix, D.~Pagani and M.~Zaro, \emph{{Large NLO corrections in
  $t\bar{t}W^{\pm}$ and $t\bar{t}t\bar{t}$ hadroproduction from supposedly
  subleading EW contributions}},
  \href{http://dx.doi.org/10.1007/JHEP02(2018)031}{\emph{JHEP} {\bfseries 02}
  (2018) 031}, [\href{https://arxiv.org/abs/1711.02116}{{\ttfamily
  1711.02116}}].

\end{thebibliography}\endgroup
\end{document}